\documentclass[final,5p,times,twocolumn,preprint,tikz, svgnames]{elsarticle}
\usepackage{amsmath}
\usepackage{graphicx}

\usepackage{cite}
\usepackage{times}
\usepackage{listings}
\usepackage{graphics}
\usepackage{latexsym}
\usepackage{amsfonts}
\usepackage{amssymb}
\usepackage{paralist}
\usepackage{xspace}
\usepackage{mathrsfs}
\usepackage{setspace}
\usepackage{color}
\usepackage{booktabs}
\usepackage{makecell}
\usepackage{wrapfig}

\usepackage{listings}

\usepackage{tikz}
\usetikzlibrary{mindmap,trees}
\usepackage{verbatim}
\usepackage{CJK}

\usepackage[colorlinks,
            linkcolor=blue,
            anchorcolor=blue,
            citecolor=blue]{hyperref} 

\allowdisplaybreaks

\newcommand{\CUTXY}[1]{{}}

\renewcommand{\paragraph}[1]{\smallskip \noindent {\textbf{#1}}}

\usepackage[linesnumbered,ruled]{algorithm2e}
\usepackage{algpseudocode}
\usepackage{amsmath}
\usepackage{multirow}

\usepackage{listings}
\usepackage{xcolor}

\usepackage{tikz}
\usepgflibrary{arrows}

\usepackage{subfig}
\newcommand{\methodname}{{\tt{AIEC}}}

\journal{Knowledge-Based Systems}

\begin{document}

\begin{frontmatter}

\title{Towards AI-Empowered Crowdsourcing}

\author[inst1,inst2]{Shipeng Wang}
\author[inst1,inst2]{Qingzhong Li\corref{cor1}}
\author[inst1,inst2]{Lizhen Cui}
\author[inst1,inst2]{Zhongmin Yan}
\author[inst1,inst2]{Yonghui Xu}
\author[inst3]{\\Zhuan Shi}
\author[inst2]{Xinping Min}
\author[inst4]{Zhiqi Shen}
\author[inst4]{Han Yu\corref{cor1}}

\affiliation[inst1]{
            organization={School of Software, Shandong University},
            city={Jinan},
            postcode={250100}, 
            country={China},
            }

\affiliation[inst2]{
            organization={Joint SDU-NTU Centre for Artificial Intelligence Research, Shandong University},
            city={Jinan},
            postcode={250100}, 
            country={China},
            }

\affiliation[inst3]{
            organization={School of Computer Science and Technology, University of Science and Technology of China},
            city={Hefei},
            postcode={230027}, 
            country={China},
            }

\affiliation[inst4]{organization={
            School of Computer Science and Engineering, Nanyang Technological University},
            city={Singapore},
            postcode={639798}, 
            country={Singapore},
            }

\cortext[cor1]{Corresponding author.\\E-mail addresses: wangshipeng95@mail.sdu.edu.cn (S. Wang), lqz@sdu.edu.cn (Q. Li), clz@sdu.edu.cn (L. Cui), yzm@sdu.edu.cn (Z. Yan), xuyonghui@sdu.edu.cn (Y. Xu), zhuan.shi@epfl.ch (Z. Shi), minxinping0105@126.com (X. Min), zqshen@ntu.edu.sg (Z. Shen),  han.yu@ntu.edu.sg (H. Yu).}

\begin{abstract}
Crowdsourcing, in which human intelligence and productivity is dynamically mobilized to tackle tasks too complex for automation alone to handle, has grown to be an important research topic and inspired new businesses (e.g., Uber, Airbnb). Over the years, crowdsourcing has morphed from providing a platform where workers and tasks can be matched up manually into one which leverages data-driven algorithmic management approaches powered by artificial intelligence (AI) to achieve increasingly sophisticated optimization objectives. In this paper, we provide a survey presenting a unique systematic overview on how AI can empower crowdsourcing to improve its efficiency - which we refer to as \textit{AI-Empowered Crowdsourcing} (\methodname{}). We propose a taxonomy which divides \methodname{} into three major areas: 1) task delegation, 2) motivating workers, and 3) quality control, focusing on the major objectives which need to be accomplished. We discuss the limitations and insights, and curate the challenges of doing research in each of these areas to highlight promising future research directions.
\end{abstract}




\end{frontmatter}


\section{Introduction}

\begin{figure}[t!]
	\centering
	\includegraphics[width=1\linewidth]{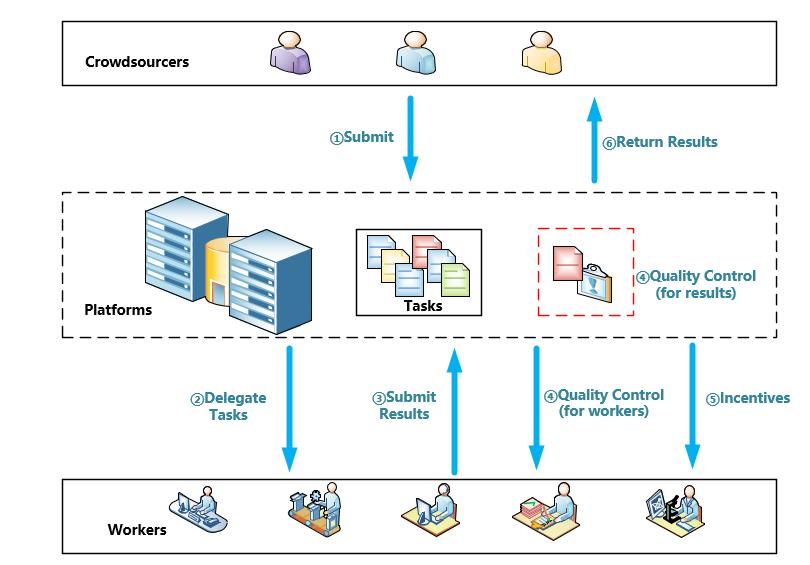}
	\caption{The key entities and main processes involved in crowdsourcing.} \label{fig:1}
\end{figure}

Crowdsourcing is an interdisciplinary approach for efficiently mobilizing the effort, time and resources of human participants to perform tasks which are relatively easy for humans but challenging for automation \citep{Doan-et-al:2011}. As illustrated in Figure \ref{fig:1}, in a typical crowdsourcing setting, \textit{crowdsourcers} (a.k.a. task requesters) propose \textit{tasks} (a.k.a. human intelligence tasks (HITs) or microtasks) and delegate them to \textit{workers} to perform through \textit{platforms}. The platforms control the task quality during and after workers undertake the tasks, and offer \textit{rewards} in exchange for workers' effort when they complete tasks. Over the years, it has inspired new applications such as Uber\footnote{https://www.uber.com/}, Airbnb\footnote{https://www.airbnb.com} and Amazon Mechanical Turk\footnote{https://www.mturk.com/} which significantly changed the way people commute, travel and work.

\subsection{About \methodname{}}
In order to achieve this design objective well, crowdsourcing systems need to solve three important problems. Firstly, it must find a efficient way to delegate the myriad tasks to suitable workers among a large pool of potential workers. Secondly, it must effectively motivate the workers to stay with the platform and put in the necessary effort to complete the tasks with good quality (e.g., through incentive engineering). Last but not least, once the results are obtained from the workers, it must verify the quality of the results in order to ensure that the crowdsourcers (who are paying for the services of the platform) are satisfied. Such large-scale and long-term ecosystem management effort is highly challenging for the human mind to accomplish, and often require algorithmic decision support at various levels.

\begin{figure}[t!]
	\centering
	\includegraphics[width=1\linewidth]{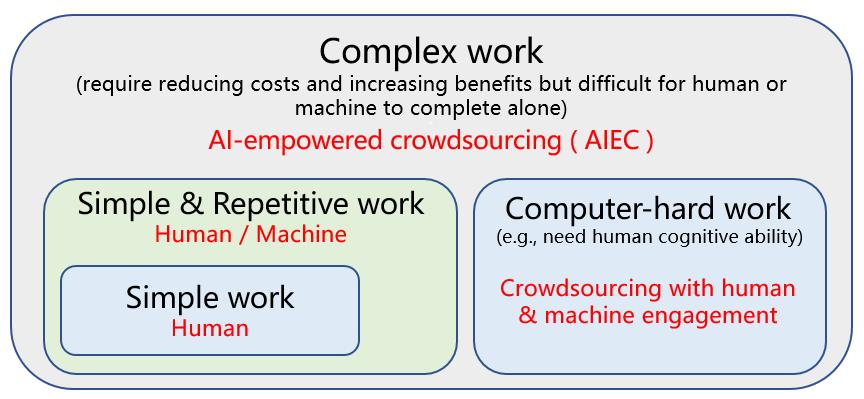}
	\caption{AI-empowered crowdsourcing (\methodname{}).} \label{fig:AIEC}
\end{figure}

The concept of algorithmic crowdsourcing was proposed by Microsoft Research\footnote{https://www.microsoft.com/en-us/research/project/algorithmic-crowdsourcing/}. It refers to technologies which aim to automate the combination of human intelligence with computing power efficiently in order to solve problems that are too difficult for either group to solve effectiveness on their own. Artificial intelligence (AI) techniques have been extensively applied to empower crowdsourcing to benefit from algorithmic management. They have been incorporated into many key stages of crowdsourcing including motivating active participation by high quality workers, disseminating tasks to them in an optimal manner, and determining which results from crowdsourcing to adopt. It has been estimated that the business sector relying on AI-empowered crowdsourcing (\methodname{}) will be worth around US\$335 billion by 2025 \citep{Lee-et-al:2015}. 
As shown in Figure \ref{fig:AIEC}, compared with traditional crowdsourcing, \methodname{} leverages artificial intelligence algorithms and tools to control the cost, quality and timeliness of crowdsourcing process, so as to reduce costs and increase benefits. It can unleash the productivity of workers and machines to a greater extent, thereby promoting efficient collaboration and ultimately improving the efficiency of crowdsourcing activities.

\subsection{Key Contributions}
There exist a number of surveys on the topic of crowdsourcing from various perspectives. Yuen et al. \citep{Yuen-et-al:2011} is a widely cited paper that offers a review on the general topic of crowdsourcing during its early stage of development. The topic has been revisited almost a decade later in \citep{Bhatti-et-al:2020}. In \citep{Jiang-et-al:2018}, a survey and tutorial on conducting crowdsourcing from the lens of multi-agent systems is provided. Other surveys focus on important steps involved in crowdsourcing. For instance, \citep{Allahbakhsh-et-al:2013IEEEInternetComputing} surveyed the issue of quality control in crowdsourcing, \citep{Muldoon-et-al:2018} reviewed incentive engineering techniques in crowdsourcing, and \citep{Hettiachchi-et-al:2022} focused on the task allocation step in crowdsourcing. Apart from surveying the components making up typical crowdsourcing systems, surveys on the application of crowdsourcing in other application domains are also available. For example, \citep{Guo-et-al:2014survey} surveyed the usage of crowdsourcing for data mining, whereas \citep{Zhou-et-al:2018survey} focused on crowdsourcing-based indoor localization.
However, there lacks a survey of \methodname{} techniques which are playing an increasingly important role in the operation of large-scale crowdsourcing platforms. 

In this paper, we provide a comprehensive survey on the state of the art of \methodname{} to bridge this gap. Our main contributions are summarized as follows:
\begin{enumerate}
    \item We propose a hierarchical taxonomy that first divides the \methodname{} literature into three major categories based on the most important functionalities require by crowdsourcing platforms: 1) task delegation, 2) motivating workers, and 3) quality control.
    \item For each of these major categories, we further divide the taxonomy according to the different approaches taken to address the technical challenges, highlighting their main ideas, assumptions made which could introduce potential limitations as well as commonly adopted performance evaluation methods.
    \item We envision promising opportunities of future research which can optimal complex process control, trusted open collaboration, transparent interactions and standardized benchmarking into the field of crowdsourcing towards making \methodname{} a reality.
\end{enumerate}
With this survey, we aim to provide a comprehensive and succinct roadmap for researcher and practitioners who are interested in entering this interdisciplinary field to conveniently gain an overview of the key advances to date, as well as promising directions of development moving forward.

\subsection{Overview of the \methodname{} Taxonomy}
\methodname{} approaches generally aim to balance the objectives of the two main stakeholders involved in typical crowdsourcing systems:
\begin{enumerate}
	\item \textit{Crowdsourcers}, whose aim is generally to maximize the quality of results obtained and minimize the latency time subject to a limited budget.
	\item \textit{Workers}, who are heterogeneous in terms of motivation to work, skill level, productivity and availability, aim to maximize the rewards they receive from completing tasks. Workers may be strategic in terms of how to price and spend effort their effort.
\end{enumerate}
Crowdsourcing tasks are often associated with deadlines before which results must be obtained in order for workers to get the rewards. in order to ensure that workers are rewarded.Models of complex tasks also take into account the skillsets required from workers.
	
\begin{figure}[t!]
	\centering
	\includegraphics[width=1\columnwidth]{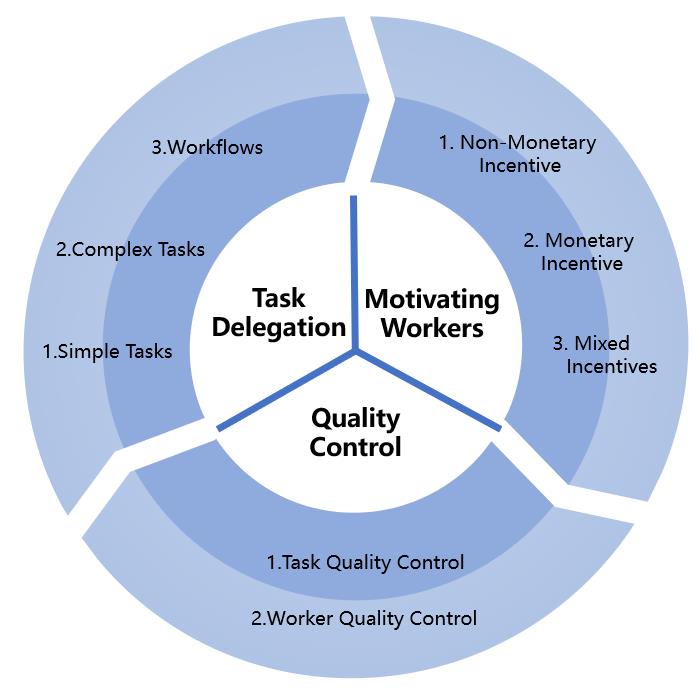}
	\caption{The proposed \methodname{} taxonomy.} \label{fig:2}
	
\end{figure}

With these considerations, we propose an \methodname{} taxonomy from the perspective of the crowdsourcing process as shown in Figure \ref{fig:2}. It first summarizes and compares the representative \methodname{} approaches designed for task delegation under different conditions (Section \ref{Task Delegation}), which often takes place at the beginning of the crowdsourcing process. Then, incentive engineering research for motivating workers either with or without monetary payout (Section \ref{Motivating Workers}), which is the driving force for the successful execution of crowdsourcing tasks, are discussed. Lastly, commonly adopted approaches for quality control from the perspective of task result quality control and worker quality control (Section \ref{Quality Control}), which ensure that the crowdsourcing process can proceed sustainably and efficiently, are presented. In the following sections, we delve into the details of each category of this taxonomy to provide an in-depth survey of key techniques.

\section{Task Delegation} \label{Task Delegation}

In typical crowdsourcing systems, tasks need to be assigned to multiple workers to be completed independently or collaboratively, and then aggregated to obtain the final results or answers. Due to the complexity of crowdsourcing tasks and the uncertainties about the workers, how to assign or delegate tasks to the appropriate workers is one of the most difficult and fundamental challenges in crowdsourcing. Task delegation algorithms aim to match tasks to workers in order to obtain high quality task outcomes, while satisfying temporal constraints, budget constraints, and sometimes other special constraints (e.g., spatial constraints in mobile crowdsourcing).

Existing task delegation algorithms in \methodname{} research typically follow one or more of the following general assumptions:
\begin{enumerate}
\item \textit{Static scenarios}: Most existing task delegation research assumes that the information of all tasks and workers are fully known before task delegation.
\item \textit{Information Unambiquity}: The task information is assumed to be unambiguous, containing at least the task requirements, task budget and task deadline, etc.
\item \textit{Crowd intelligence}: a large number of unknown workers with various expertise and skills can replace domain experts.
\end{enumerate}

\begin{figure*}[t!]
\centering
\begin{tikzpicture} 
[
    mindmap, every node/.style=concept, concept color=orange, text=white, scale = 0.9, 
    level 1/.append style={level distance=4.5cm, sibling angle=120, font=\large},
    level 2/.append style={level distance=3.5cm, sibling angle=60},
    level 3/.append style={level distance=3cm, sibling angle=60}
  ]
  \node{\textbf{\huge{Task \\ Delegation}}} [clockwise from=90]
  child [concept color=Orchid] {
      node {Simple Tasks} [clockwise from=50]
      child { node {Static/ Offline Delegation  \citep{Cavallo-Jain:2012, Heidari-Kearns:2013, Jain-et-al:2014, Jain-et-al:2016AAMAS, Biswas-et-al:2015AAMAS, Liao:2021ESA, Tu:2020TKDD} } 
            }
      child { node {Dynamic/ Online Delegation \citep{Yu-et-al:2013IJCAI, Yu-et-al:2017SciRep, Yu-et-al:2014AAMAS, Yu-et-al:2017ICA,Yu-et-al:2019AIES, Chandra-et-al:2015, Chen-et-al:2014, Goel-et-al:2014, Assadi-et-al:2015, Li:2021ComputerJournal} } }
    }
  child [concept color=SkyBlue] {
      node {Complex Tasks} [counterclockwise from=290]
      child { node {Worker Ability Pre-assessment \citep{Bragg-et-al:2014, Tang:2020IJCAI, Cui:2017IJCS}}}
      child { node {Multi-stage Delegation with Evaluation \citep{Wang-et-al:2017, Rangi-et-al:2018} }}
      child { node {Optimization of Worker Combinations \citep{Pan-et-al:2016, Abououf:2019JNCA, Li-et-al:2021ICDE, Samanta:2021GLOBECOM, Kang:2020TSC, Yin:2021PTPNA} }}
    }
  child [concept color=Lavender] {
      node {Optimizing Workflows} [counterclockwise from=120]
      child { node {Hierarchical Delegated Subtasks \citep{Nath-Narayanaswamy:2014, Kamar-Horvitz:2015, Yu-et-al:2015,Yu-et-al:2016SciRep} }}
      child { node {Interdepend-ent Subtasks \citep{Yang:2019SmartWorld, Tran-Thanh-et-al:2015, Ni2020icde, Liu-et-al:2022IS} }}
      child { node {Independent Subtasks \citep{Charles-et-al:2015, Goto-et-al:2016}  }}
    };
\end{tikzpicture}

	\caption{The \methodname{} taxonomy on task delegation in crowdsourcing systems.} \label{fig:TaskDelegation}
\end{figure*}
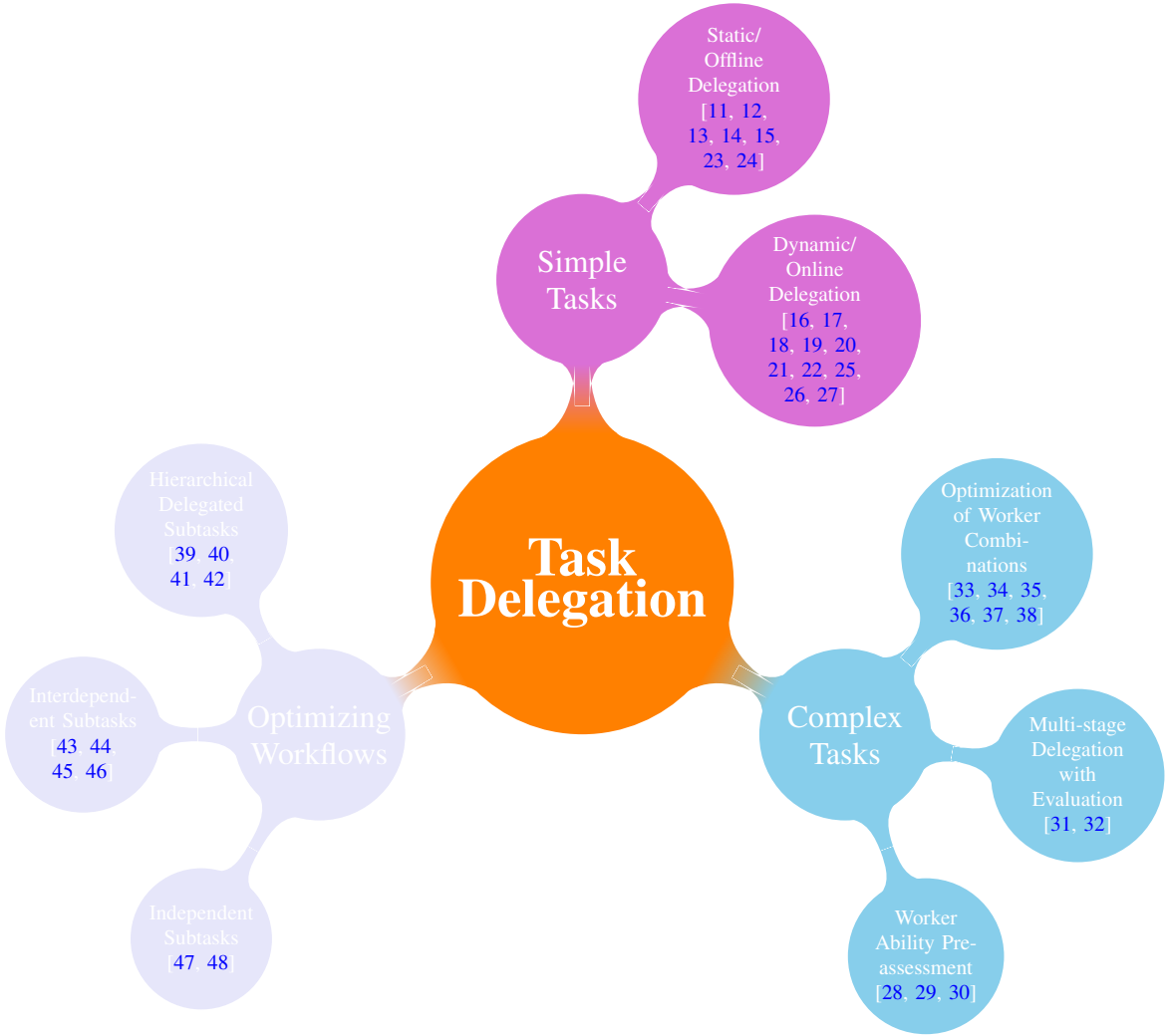

Current task delegation approaches generally combine workload-balancing among workers with considerations for their respective skills, productivity, and availability in an effort to make quality-time-cost trade-offs to achieve diverse design objectives. They can be further divided into three categories according to how they treat tasks: 1) delegating simple tasks which can be completed by one worker (Section \ref{Delegating Simple Tasks}); 2) delegating complex tasks which require collaboration by workers with complementary skills (Section \ref{Delegating Complex Tasks}); and 3) optimizing workflows consisting of many interrelated or sequentially dependent simple subtasks (Section \ref{Optimizing Workflows}). The proposed \methodname{} taxonomy on task delegation is illustrated in Figure \ref{fig:TaskDelegation}.

\subsection{Delegating Simple Tasks} \label{Delegating Simple Tasks}
\textit{Simple Tasks} are tasks that can be carried out by a single worker and do not require workers to collaborate on the tasks in the process. Individual workers can utilise their skills to complete simple tasks. A series of research focusing on the simplest setting of delegating homogeneous subtasks have emerged over the years, which mainly include game theory, queuing systems, auction theory, multi-armed bandit approach and other theories, all of which are dedicated to optimising the solution of simple task allocation problems.

The assignment of simple tasks to individual workers was initially explored by focusing on offline algorithms.
In \citep{Cavallo-Jain:2012}, the authors proposed an approach which combines decision theoretic optimization and game theory for the purpose of maximizing the quality of the results with a limited budget when allocating contest-based crowdsourcing tasks to workers. In \citep{Heidari-Kearns:2013}, the authors proposed a decision theoretic optimization approach to effectively organize a population of workers with varying capabilities to balance the workload among them.

The varying nature of the workers' qualities and costs makes task allocation a non-trivial problem in almost all crowdsourcing applications. A multi-armed bandit approach \citep{Jain-et-al:2014} was proposed which strategically selects a subset of workers with unknown qualities for task allocation in order to achieve a desired level of accuracy for binary labelling tasks. Subsequently, they designed a novel task allocation mechanism that allocates blocks of tasks to a worker instead of a single task in \citep{Jain-et-al:2016AAMAS}, which can reduce the overhead of frequent assignment of low-cost tasks, while ensuring the stability of the workers on the crowdsourcing platform to a certain extent.
In \citep{Biswas-et-al:2015AAMAS}, the authors further assumed that workers are strategic about their costs, and proposed an interval cover mechanism which leverages the natural linear ordering structure in some crowdsourcing applications to compute optimal task allocation plans.

Although offline simple task delegation algorithms can achieve a balance between task cost, time and quality of task outcome, they are in most cases unable to cope with real-time arrival of task demand and constant worker variation, which requires real-time online algorithms. 

In \citep{Yu-et-al:2013IJCAI,Yu-et-al:2017SciRep}, the authors adopted ideas from queuing systems and treated workers with limited productivity as first-come-first-served task queues. By leveraging the structure of the problem, distributed and centralized Lyapunov optimization-based approaches were proposed which recommend the number and types of new tasks to workers in real time in order to maximize time-averaged system utility while minimizing latency. A subsequent work \citep{Yu-et-al:2014AAMAS} adjusted the utility function for situations in which worker utility is non-linearly related to the number of tasks they complete over time (i.e. workers' preference for rest increases as they complete more tasks over time). The extensions of this work opportunistically schedule how workers' complete the tasks in order to achieve work-life balance while preserving system-level task throughput \citep{Yu-et-al:2017ICA,Yu-et-al:2019AIES}.

In \citep{Chandra-et-al:2015}, the authors imposed an additional constraint on the task allocation problem - unreliable and strategic workers arrive over time into the system. They proposed a dynamic pricing mechanism and an auction-based mechanism to jointly solve the problem of task delegation and payment decisions. In \citep{Chen-et-al:2014}, a heuristic algorithm was proposed to sequence mobile crowdsourcing tasks by taking into account workers' current trajectories in order to minimize disruptions to their movement.

A number of works focused on delegating heterogeneous tasks each requiring a specific, possibly different, skill. In \citep{Goel-et-al:2014}, the authors treated the problem of matching workers with different skills and interests with heterogeneous tasks as a bipartite graph. They designed a task allocation approach which ensures budget feasibility, incentive-compatibility, and achieves near-optimal utility. In \citep{Assadi-et-al:2015}, the authors approached this problem from the perspective of the crowdsourcers who has a collection of heterogeneous tasks to be delegated. Workers are assumed to arrive one-by-one and each declares a set of tasks he can solve and desired payments. An online task allocation algorithm was proposed with a provable upper bound on its competitiveness over an arbitrary sequence of workers who want small payments relative to the crowdsourcer's total budget.

To maximize task quality within a limited budget, \citep{Tu:2020TKDD} proposed a model which jointly models the worker's bias, the ground truth task results and the task features, and dynamically assigns tasks to more appropriate groups of workers by creating a mapping between worker annotations and task features. In \citep{Liao:2021ESA}, the authors proposed an algorithm to recommend a worker group based on multi-community collaboration. It combines worker reputation, preference and activity characteristics to divide the workers into groups based on the similarity of behaviour patterns.

The simple task delegation algorithms over the social network could disseminate task information by means of the word-of-mouth pattern. In \citep{Li:2021ComputerJournal}, the authors defined the CBTA problem, which aims to make task information available to as many target social network groups as possible in crowdsourcing to ensure the task delegation is cost-effective as well as budget-balanced. Meanwhile, the authors proposed CB-greedy algorithm based on the greedy strategy and CB-local algorithm based on the local search technique, which can ensure that tasks can be effectively delegated to a greater number of target workers. CB-local algorithm is especially efficient for reducing the overhead of task delegation.

\subsection{Delegating Complex Tasks} \label{Delegating Complex Tasks}
Automatically delegating \textit{Complex Tasks} which require workers with different skills to collaborate remains a challenging and worthwhile area to explore in \methodname{}. Algorithms for assigning complex tasks can currently be divided into three levels: pre-assessment of worker capabilities, assessment of worker capabilities for reassignment during task execution, and optimization of the worker combination for the task.

Pre-evaluating worker capabilities before delegating the complex task can reduce the cost overhead of a task while accurately delegating it, and avoid task delays or failures due to unreasonable task delegation, which is the starting point for most research on complex task delegation algorithms. In \citep{Bragg-et-al:2014}, the authors proved that finding the best allocation of tasks with diverse difficulties to workers with varying skills is NP-hard. They further developed a set of approximation algorithms which attempt to route the tasks to the most appropriate workers.

For complex tasks that require multiple skills, a weighted multi-skill tree (WMST) model was proposed in \citep{Tang:2020IJCAI} to quantitatively analyze multiple skills and correlations among them. On the basis of WMST, an adaptive acceptance-expectation-based task assignment algorithm was proposed to delegate tasks to the most suitable workers even unavailable temporarily. This algorithm is effective in macrotask-oriented service crowdsourcing systems which require multiple professional skills and have a relatively stable and predictable pool of professional workers.
While taking quality-time-cost trade-offs into account, crowdsourcing tasks are more likely to be delegated to workers who have the time to complete them, avoiding the herding effect due to a backlog of tasks for certain workers. Several algorithms to avoid the herding effect have been proposed in \citep{Yu-et-al:2013IJCAI,Yu-et-al:2017SciRep}, which are effective for the delegation of tasks that require simple or single-skilled tasks. In \citep{Cui:2017IJCS}, the authors combined supervised learning and reinforcement learning to assign tasks to workers. Making the most of policy networks as well as reputation networks, the proposed method can explore for better task allocation strategy while ensuring the prediction of trends in worker reputation fluctuations.

Furthermore, in order to adapt to the dynamics of the task while adjusting to unreasonable worker delegation schemes, some researches have focused on further evaluating worker capabilities during task execution and thus optimally generating new delegation strategies.
A number of efficient algorithms for delegating complex tasks have been proposed based on the estimation of task difficulty and worker expertise. In \citep{Wang-et-al:2017}, the authors proposed a two-stage task delegation algorithm. In the first stage, it exposes a small portion of sample tasks to multiple workers to estimate the difficulty of the tasks. Then, a model with this training set is built to predict the difficulty of new tasks in the second stage. This method can distinguish between easy and hard tasks before delegating them to workers with different capabilities to improve result quality. 
In \citep{Rangi-et-al:2018}, the authors developed the notion of Limited-information crowdsourcing systems, where crowdsourcers can allocate tasks according to the knowledge about workers' capabilities acquired over time. They modelled the task delegation problem as a multi-armed bandit problem that is arm-limited and budget-limited, and implemented an efficient worker selection policy by means of the bounded KUBE(B-KUBE) algorithm. To estimate the workers' capabilities, this method takes workers' value contributions for inference rule into account, and then clearly illustrates the relation between the classification errors and the workers' cumulative value contributions.

Different combinations of workers may affect the quality and efficiency of task completion, especially many complex crowdsourcing tasks are similar or related. For example, some mobile crowdsourcing tasks may be geographically relevant. A proper selection of workers when delegating such tasks can improve crowdsourcing effectiveness.

The approach in \citep{Pan-et-al:2016} is the most well documented approach for dynamically forming teams of workers with appropriate combinations of skills to tackle complex crowdsourcing tasks. The authors considered the challenge of ensuring efficient quality-time-cost trade-offs in collaborative crowdsourcing in which a task requires a combination of diverse skills from different workers to complete. They proposed an optimization-based algorithm which automatically determines how to make full use of the expertise of candidate workers to maximize the quality of results, as well as minimize latency and staying budget-balanced. Analysis shows that the proposed approach, if adhered to by all workers, can achieve close to optimal collective utility. In \citep{Abououf:2019JNCA}, the authors developed a multi-worker multi-task selection framework. It first proposed a clustering method based on k-mediods to cluster the tasks according to their geographical proximity. Then, it assigns a cluster of workers to a group of tasks using genetic algorithm. Finally, they employed Tabu search algorithm to delegate workers to every task in a cluster according to the task completion time.

When the task becomes more complex, it may be impossible for a single worker to complete a task. This has led to research into the problem of group task delegation. In \citep{Li-et-al:2021ICDE}, the authors developed a preference-aware group task delegation framework, including mutual information-based preference modeling and preference-aware group task assignment. The former is responsible for the preference learning of groups of workers based on interaction data, and the latter is used to delegate tasks to proper worker groups using a tree-decomposition algorithm, which simultaneously maximises the number of task assignments and workers' preferences. Following the same intuition, \citep{Samanta:2021GLOBECOM} considered workers' willingness for the delegation of complex tasks to reduce the failure rate of task assignment. The authors proposed a skill-oriented dynamic task delegation method based on the greedy algorithm, which incorporates workers' willingness. This method iteratively delegates workers to tasks based on the skills that are desirable to perform the tasks and the workers' skillsets. In addition, A multi-armed bandit (MAB) framework is devised in \citep{Kang:2020TSC} to learn a worker's preferences, which is employed to ensure that workers assigned to a task have a higher probability of accepting it and completing it. Further, Deep Q-Network is applied to task allocation in dynamic worker and task situations to maximize the benefits of workers and task requesters in \citep{Shan:2020ICDE}.
In addition, for a set of heterogeneous crowdsourcing tasks with multiple collaborators, \citep{Yin:2021PTPNA} formulated the heterogeneous task delegation problem as an optimisation problem and modelled the task delegation as a hidden Markov process. The authors utilized the Viterbi algorithm to explore efficient task assignment strategies that maximize the total value of tasks received by workers.

\subsection{Optimizing Workflows} \label{Optimizing Workflows}
Some crowdsourcing tasks can be divided into multiple subtasks that have temporal relationships or are interrelated or independent, and these subtasks naturally give rise to workflows due to the effects and relationships between them. There are currently many algorithms that aim at \textit{Optimizing Workflows} to improve the efficiency and profitability of crowdsourcing tasks, mainly focusing on Hierarchical Delegated Subtasks, Interdependent Subtasks and Independent Subtasks. 

Hierarchical crowdsourcing in which tasks pass through multiple workers in a network of trust naturally form dynamic workflows. In \citep{Nath-Narayanaswamy:2014}, the authors designed a game theoretic approach to determine how to divide the workers' effort between working on the tasks and finding other suitable workers to pass along the tasks in order to achieve improved overall efficiency while recruiting new workers. In \citep{Kamar-Horvitz:2015}, the authors focused on hierarchical consensus tasks which aim to find right answers by means of a hierarchy of subtasks. They constructed multiple hierarchical classification models that combine machine and human wisdom, and used Monte Carlo planning to constrain the policy space using task structure in order to improve tractability. In \citep{Yu-et-al:2015,Yu-et-al:2016SciRep}, optimization-based approaches were proposed to jointly determine how many new tasks a worker should accept, how many existing tasks a worker should sub-delegate to others (and to whom), as well as how to price a worker's effort based on the current context, in order to maximize the quality of the task results while minimizing latency.

The interdependencies among the subtasks of some crowdsourcing tasks can be very tight. If crowdsourcing tasks with internal dependencies are not delegated properly, it will not only lead to delays and budget overrun, but may even lead to task failure. Therefore, subtasks cannot simply be decomposed and delegated to workers from the source task following a partitioning approach. To avoid these problems, the idea of dividing tasks into logically related subtasks and then ranking the subtasks using a depth-first search algorithm was proposed in \citep{Yang:2019SmartWorld}. The authors proposed a dynamic allocation method, SATD, which first divides the subtasks into serial tasks. Then, it takes the capabilities and budgets of workers into account in the context of subtask ranking and uses the greedy approach to assign the subtasks to the reliable workers with the lowest budgets, while ensuring that each worker is able to complete as many tasks as possible within a given duration. In \citep{Tran-Thanh-et-al:2015}, the authors focused on task allocation with multiple complex workflows, each consisting of multiple inter-dependent tasks. They proposed an algorithm which first calculates an efficient allocation of a limited budget to each workflow, and then determines the number of and prices for the tasks within each workflow.

In \citep{Ni2020icde}, the authors defined dependency-aware problem in spatial crowdsourcing and proposed greedy approach and game-theoretic algorithm to optimize subtasks delegation with dependency. The greedy approach greedily allocates a group of tasks with the largest size to workers, then the game-theoretic algorithm further ensures that more tasks are assigned to workers. 
Likewise, \citep{Liu-et-al:2022IS} solved the complex task assignment problem in spatial crowdsourcing, which is dedicated to assigning multi-stage complex tasks to workers and maximising total profit even under multiple constraints such as dependency, skills and budget. They assigned the most profitable workers to subtasks by employing a greedy algorithm and guided the worker to adjust individual strategy in response to the strategies of others through a game algorithm.

In addition to the two categories mentioned above, there are also complex tasks that can be divided into multiple independent subtasks that have potential serial or parallel workflows and also need to be delegated appropriately to save budget and improve quality.
In \citep{Charles-et-al:2015}, a graphical framework was proposed to describe and monitor crowdsourcing workflows. The framework can be representative of the workflows of most of the widely known crowdsourcing applications, and formally define adaptive workflows which depend on workers' skills and/or task deadlines. In addition, it allows the expression of constraints on workers based on workers' contributions. 
Nevertheless, even with these parallel and iterative workflow optimization approaches, the characteristics of workflow in relation to tasks and the environment are not fully understood. To bridge this gap, \citep{Goto-et-al:2016} proposed a workflow model jointly considering workers' abilities, the difficulty to improve task results, and the preference of the crowdsourcers. The authors showed that the optimal workflow can be found using a search algorithm.

\section{Motivating Workers} \label{Motivating Workers}


Motivating workers to perform tasks or subtasks is an important topic in crowdsourcing research. Crowdsourcers design incentive mechanisms to encourage desirable behaviours from workers. The general goal is to maximize the participation of workers for the purpose of maximizing the utility of the crowdsourcing system or crowdsourcers under the overarching assumption that workers are rational and self-interested. 


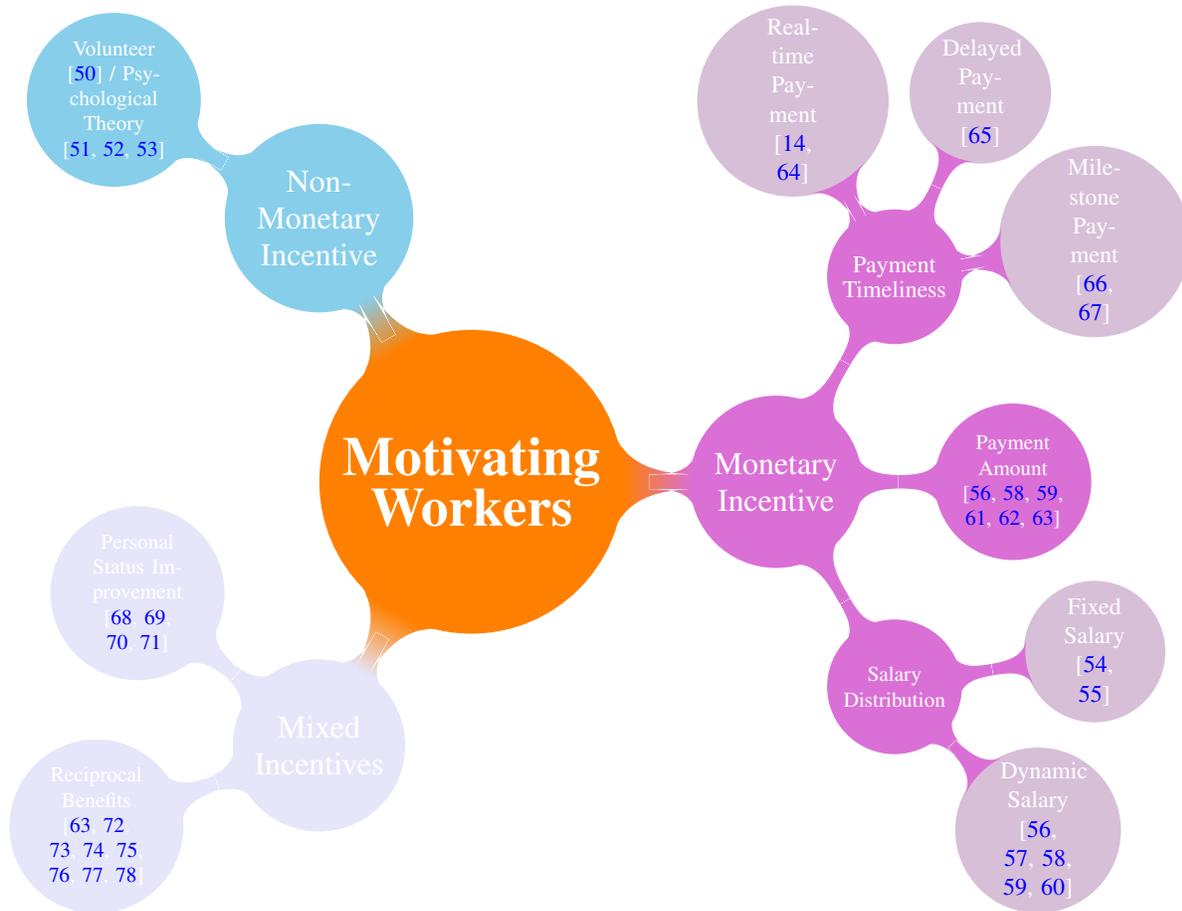
\begin{figure*}[t!]
\centering
\begin{tikzpicture} 
[
    mindmap, every node/.style=concept, concept color=orange, text=white, scale = 0.9, 
    level 1/.append style={level distance=4.5cm, sibling angle=120, font=\large},
    level 2/.append style={level distance=3.5cm, sibling angle=60},
    level 3/.append style={level distance=3cm, sibling angle=55, font=\small}
  ]
  \node{\textbf{\huge{Motivating \\ Workers}}} [clockwise from=120]
  child [concept color=SkyBlue] {
      node {Non-Monetary Incentive} [clockwise from=150]
      child { node {Volunteer \citep{Dittus-et-al:2016CSCW}  / Psychological Theory \citep{Xue-et-al:2013, Elmalech-Grosz:2017, Segal-et-al:2016}  } 
            }
    }
  child [concept color=Orchid] {
      node {Monetary Incentive} [counterclockwise from=300]
      child { node {Salary Distribution }[clockwise from=10]
                  child {node[concept,color=Thistle,text=white] {Fixed Salary \citep{Yu-et-al:2015AAMAS, Hu-Zhang:2017} }} 
                  child {node[concept,color=Thistle, text=white] {Dynamic Salary \citep{Wang-et-al:2016CN, Wei-et-al:2015INFOCOM, Liu-et-al:2022IEEEIoT, Biswas-et-al:2015, Lu-et-al:2018ACMCEC} }}
      }
      child { node {Payment Amount \citep{ Wang-et-al:2016CN,  Liu-et-al:2022IEEEIoT, Biswas-et-al:2015, Radanovic-Faltings:2016, Fang-et-al:2017WWW, Nie-et-al:2018GLOBECOM}  }}
      child { node { \small{Payment Timeliness} }[clockwise from=120]
                  child {node[concept,color=Thistle,text=white] {\small{Real-time Payment \citep{Jain-et-al:2016AAMAS, Liu-et-al:2020DASFAA} }}} 
                  child {node[concept,color=Thistle,text=white] {\small{Delayed Payment \citep{Chandra-et-al:2015AAAI} }}}
                  child {node[concept,color=Thistle,text=white] {\small{Mile-stone Payment \citep{Difallah-et-al:2014, Yin-et-al:2015} }}}
      }
    }
  child [concept color=Lavender] {
      node {Mixed Incentives} [counterclockwise from=140]
      child { node {Personal Status Improvement \citep{Zhang-et-al:2012INFOCOM, Rui-et-al:2016JEIT, Miao-et-al:2016, Wei-et-al:2020Electronics} }}
      child { node {Reciprocal Benefits \citep{ Nie-et-al:2018GLOBECOM, Lv-Moscibroda:2015, Gong-et-al:2015TSIPN, Nie-et-al:2018TWC, Nie-et-al:2020TWC, Wang:2020TMC, Xu:2022TMC, Huang:2020TMC}  }}
    };
\end{tikzpicture}
	\caption{The \methodname{} taxonomy for motivating workers.} \label{fig:motivation}
\end{figure*}

To understand the background of existing incentive mechanism design approaches in \methodname{}, it is crucial to examine the assumptions made. The following assumptions are essential and commonly adopted in the field of \methodname{} when devising incentive mechanism:
\begin{enumerate}
\item \textit{Self-interested Crowdsourcers}: The goal of the crowdsourcer is to maximize own benefit or utility. 
\item \textit{Self-interested and Rational Workers}: To say that an worker is self-interested  and rational is to say that the worker will perform in accordance with individual interests or goals. The worker will always move towards the goal of maximising his or her own financial benefits, regardless of the strategies of competitors.
\item \textit{Adequate Worker Pool}: The vast majority of existing incentive studies for workers assume that there are enough participants to perform the tasks.
\item \textit{Limited Budget}: Crowdsourcers have limited budgets for incentives that involve monetary.
\end{enumerate}

Based on the above common assumptions, different reviews and surveys have already been conducted on motivating workers in terms of incentive mechanisms and techniques in various topics of crowdsourcing. In \citep{Katmada-et-al:2016ICIS}, the authors provided a holistic overview of the incentive mechanisms used in crowdsourcing environments, including four main directions: reputation schemes, gamification practices, social mechanisms, and financial rewards and career opportunities. 
However, this paper is only an overview of user motivation and incentive mechanisms, and does not focus on the design and use of specific incentive mechanisms. 
In \citep{Muldoon-et-al:2018KER}, the authors addressed crowdsourcing incentives across multiple disciplines, such as incentives in game theory and behavioural economics. It specifically pays attention to the problem in the crowdsourcing systems  based on game theoretic approaches and attempts to push these approaches to broader researches and applications. Nevertheless, this survey focuses more on monetary incentives and explores the application of game theory-based incentive mechanisms to commercial platforms to improve the utility and benefit of crowdsourcers or platforms. 
It overlooked the role of non-monetary incentive mechanism in AI-powered crowdsourcing. 

There are also works that summarize the design of incentive mechanisms at the algorithmic level, such as \citep{Tong-et-al:2020VLDBJournal}. It identified four key algorithmic domains in spatial crowdsourcing, of which the design of the incentive mechanism is a crucial part. In terms of incentive mechanism design in spatial crowdsourcing, the authors divided existing incentive mechanisms into two types: 1) posted price models and 2) auction-based models. Similar to the previous work \citep{Muldoon-et-al:2018KER}, this survey does not discuss non-monetary incentive mechanisms such as volunteer-based crowdsourcing and social participation. At the same time, the incentive mechanism of spatial crowdsourcing may not be directly extended to other crowdsourcing scenarios.
There is also literature focusing on how to motivate workers from the perspective of crowdsourcing process technology.

In this part of our survey, we set out to complement existing surveys on crowdsourcing incentive mechanism design to address the aforementioned limitations.
In general, three broad categories of approaches have been devised to motivate crowdsourcing workers: 1) volunteer-based approaches which involve no monetary incentive (Section \ref{Motivating Workers without Payment}); 2) payment-based approaches which mainly involve the use of monetary incentives (Section \ref{Motivating Workers with Payment}); and 3) mixed incentives approaches which integrate monetary incentives with other social incentives (Section \ref{Motivating Workers with Mixed Incentives}), such as reputation and reciprocal service. The majority of current research belongs to the payment-based approaches category and mixed incentives approaches category. In later parts of this section, we discuss the key insights and methods of representative \methodname{} approaches belonging to each of these three categories. Before that, we highlight notable empirical studies (Section \ref{Empirical Studies}) which laid the foundation of incentive mechanism design in \methodname{}. The proposed taxonomy of the algorithms on motivating workers is shown in Figure \ref{fig:motivation}.

	

\subsection{Empirical Studies}  \label{Empirical Studies}
In order to design optimal incentive mechanisms that can achieve the goals of motivating and retaining workers, it is necessary to understand the dynamics of crowdsourcing and workers' responses to various incentives in practice. Since 2013, a series of empirical studies have been conducted seeking to answer these research questions.

In \citep{Araujo:2013}, the authors found, through a quantitative study on the 99designs crowdsourcing platform, that higher financial incentives do not always increase the quality of outcomes. Instead, the probability of obtaining successful solutions depends on the number of effective workers involved. Approaches to measure the effective retention of crowd workers have also been empirically studied in \citep{Jacques-Kristensson:2013}. The authors developed three intuitive metrics to monitor workers' willingness to work under different conditions: 1) \textit{conversion rate}, a scaled measure of effective workers who completed the task, allowing comparisons of the intrinsic factors of tasks; 2) \textit{conversion rate over time}, the cumulative conversion rate over time, providing a way to study external factors; and 3) \textit{nominal conversion rate}, the average conversion rate for an inter-quartile time period of the lifespan of a task, compensating the high variability at the beginning and the end of a task. Through experiments on Amazon's Mechanical Turk (mTurk), the authors concluded that a clear value proposition in the preview statement of a task makes it more attractive to workers, and that controlling the conversion rate helps to retain more high quality workers.

A further study on the pricing schemes for worker retention has been conducted in \citep{Difallah-et-al:2014}, through which they found that the punctuality of rewards at each milestone affects the workers' willingness to stick to their batch of tasks. Besides, the study shows that the effectiveness of pricing schemes depends on the task types w.r.t. the length of tasks, and the initial training required. For the tasks that require initial training, paying workers for their learning improved retention rates and work quality even if payments were reduced after the training phase.
Another study \citep{Ikeda-et-al:2016CHI} tested the incentives of worker retention introduced by behavioral economics and psychology. Specifically, the authors evaluated that a clear goal setting increased task completion rate because people tend to achieve that goals once they have invested efforts or funds. In addition, the study showed that offering material goods or coupons produced negative effect on participation and task completion rates due to motivational differences among individuals as implied in the post-study survey.

Apart from monetary incentives, workers can be also inspired by other factors such as reputation, altruism and fun. In \citep{Borromeo-et-al:2016}, the authors found through a comparative experiment, that volunteers completed tasks slower but produced results with similar or higher quality compared to paid workers. Another study \citep{Baruch-et-al:2016} corroborated the findings that volunteers are more interested in the quality and impact of their work. Retaining volunteer workers depends strongly on whether the platform can provide feedbacks and effective interactions for them. 
Besides, it has been proved that social relations can motivate workers in \citep{Kobayashi-et-al:2015CSCW}. This finding could provide guidance for building crowdsourcing communities for social welfare goals, such as supporting public social services and helping the disabled.



\subsection{Motivating Workers without Payment} \label{Motivating Workers without Payment}
Volunteer-based crowdsourcing depends critically on engaging workers and maintaining their interest in the tasks over time. Approaches to motivate volunteers are often closely coupled with the nature of tasks to produce intrinsic incentives (i.e. sense of achievement and satisfaction) for workers.

In \citep{Xue-et-al:2013}, the authors designed a probabilistic maximum coverage-based optimization approach to increase the probability for citizen science workers engaged in bird watching to encounter various bird species based on their given locations. The approach trades off the informativeness of the results produced by the workers with improved chances for them to accomplish the tasks. This enhances workers' sense of achievement, thereby retaining their interest.

Other than satisfying workers' sense of achievement, arousing their sense of responsibility can also improve engagement. In \citep{Elmalech-Grosz:2017}, the authors focused on crowdsourcing tasks for which ground truth information is not available (e.g., asking workers about their opinions on some issues). They designed three approaches to encourage workers' engagement for non-ground truth tasks based on the psychological theory of commitment. Workers commit to the tasks by 1) signing a contract; 2) listening to a recording; and 3) recording a personal commitment. It was found that methods 2 and 3 are more effective in engaging worker.

Continually providing volunteer workers with intrinsic incentives to maintain their engagement may not be enough. In \citep{Segal-et-al:2016}, the authors proposed a methodology for enhancing worker engagement through a combination of machine learning and extrinsic intervention design. The methodology uses real-time predictions about future worker engagement for guiding interventions, where the messages based on the proximity to the predicted worker disengagement time are delivered to workers. The results indicate that messages highlighting the helpfulness of individual workers, when delivered according to predicted times of disengagement, significantly increased their contributions. This effect was not observed when such messages were delivered at random times.



Up to now, existing work on volunteer-based incentives has mainly focused on empirical analysis and discussion, as well as designing the supporting platforms and tools to recruiting volunteers\citep{Dittus-et-al:2016CSCW}, instead of designing algorithms based on incentive theory and mechanisms. Purely volunteer-based crowdsourcing work usually relies on the intrinsic motivation of the task, allowing people to obtain more emotional satisfaction and personal improvement. Since it is difficult to quantify emotional factors such as personal achievement and satisfaction and these emotional factors are not stable enough, the crowdsourcing task relying on these non-monetary incentive mechanisms is only suitable for specific scenarios (e.g., public welfare crowdsourcing platforms and public service crowdsourcing tasks).

Although non-monetary incentives can motivate workers to complete tasks to some extent, the productivity of non-monetary driven crowdsourcing is limited. Most crowdsourcing algorithms are therefore still inseparably driven by monetary incentives.


\subsection{Motivating Workers with Payment} \label{Motivating Workers with Payment}

In the modern world, the incentive of payment is able to increase worker engagement more than entertainment and recreation in most cases, which has been also illustrated in \citep{Silberman-et-al:2010XRDS}. Motivating workers with monetary rewards significantly increases workers' contribution while providing better task completion rate. Payment-based crowdsourcing is more common in AI-powered crowdsourcing research. Existing \methodname{} monetary incentive research focuses on payment distribution, payment timeliness, and payment amount.

\subsubsection{Payment Distribution} 

The monetary incentive for workers comes mainly from the salary paid to them by the crowdsourcers. From the perspective of salary distribution, motivating workers with payment can be divided into \textit{fixed salary incentives} and \textit{dynamic salary incentives}.


\textbf{Fixed Salary Incentives.} The salaries are pre-priced by the crowdsourcer or crowdsourcing platform when the task is released. The crowdsourcer decides a take-it-or-leave-it reward offer to every worker for performing the task, and the workers can only determine whether they accept the task or not. Posted-price mechanisms are widely used in spatial crowdsourcing. Crowdsourcers will formulate rewards based on the quality of workers, such as worker's reputation. In \citep{Yu-et-al:2015AAMAS}, the authors introduced reputation as a measure of worker quality, further determining the rewards for different workers. The rewards paid to workers are based on their reputation level, in other words, the higher the reputation the more rewards a worker will receive. To maximize the quality of the results with a limited total budget, the authors proposed a spatial crowdsourcing task delegation algorithm based on the quality and budget which simultaneously takes into consideration both the reputation of the workers and the proximity to the tasks locations. And the extensive numerical experiments demonstrated the superiority of the proposed method through significant reduction in the average error rates. Posted-price mechanisms are also used in microtask crowdsourcing. In \citep{Hu-Zhang:2017}, the authors proposed a posted-price mechanism which relaxes the assumption of a finite price range. They converted the pricing problem into a MAB problem and developed an optimal algorithm to take advantage of the distinct features of microtask crowdsourcing. The performance upper bound in an unknown prior price range has been shown through theoretical analysis. 

.

\textbf{Dynamic Salary Incentives.}  \label{Dynamic Salary Distribution} For some complex crowdsourcing tasks, crowdsourcers and crowdsourcing platforms may only know the budget when releasing the task, but it is difficult to estimate the exact task rewards. The \textit{dynamic salary incentives} allow workers to negotiate rewards with crowdsourcers, rewards are tailored and customized, in which monetary incentives are dynamically adjustable based on real-time status of the crowdsourcing process. Auction mechanism is often adopted in the algorithm design of dynamic salary incentives. 

In \citep{Wang-et-al:2016CN}, the authors proposed an incentive mechanism including a two-stage auction algorithm and an online reputation management algorithm. Through this incentive mechanism, the mobile crowdsourcing system allows for static selection of workers, followed by dynamic selection of winners for task delegation after bidding to improve the efficiency and utility.
Furthermore, double auctions have proven to be an effective and successful incentive paradigm for balancing market demands when there are multiple crowdsourcers and workers in the crowdsourcing market. In \citep{Wei-et-al:2015INFOCOM}, the authors built a general framework for online double auctions in dynamic mobile crowdsourcing, and  proposed a price-ranked matching rule, a payment scheme and four price schedules, including fixed price schedule, history-mean schedule, window-history-mean schedule and McAfee-based price schedule. The proposed online double auctions incentive mechanism  can achieve truthfulness and budget balance by theoretical analysis. On the basis of the aforementioned research, three requirement-based online auction models (OSS model, OSM model and OMM model)  in on-demand service crowdsourcing are designed in \citep{Liu-et-al:2022IEEEIoT} to meet diverse demands and supplies in various crowdsourcing bidding. The authors proposed an online double auction mechanism for all the models developed above on the basis of the McAfee double auction. Through extensive simulations, the results showed that the proposed mechanisms can be adapted to the diverse demands from different users.


In addition, the superiority of the dynamic salary incentive mechanism is also reflected in the fact that the crowdsourcers can dynamically adjust the salary according to the actual progress of the task, so as to improve the enthusiasm of workers and the quality of task performance. 
In \citep{Biswas-et-al:2015}, an algorithm called STOC-PISCES has been proposed which is a multi-armed bandit framework to determine the value of payments based on the importance of the task under changing conditions (e.g., traffic reports during peak hours at crucial road junctions is more important than at other places or times). This approach is able to dynamically encourage workers to focus on high-valued tasks in a generalized setting. 
Another well-known and representative algorithm of dynamic salary incentives is the surge pricing incentive mechanism. In \citep{Lu-et-al:2018ACMCEC}, the authors analyzed driver movements using a multinomial logit model over drivers direction of motion. 
As a result, the designed incentive mechanism ensures that the participants are reliable and able to complete tasks as quickly as possible. At the same time thee proposed surge pricing strategy not only save waiting time for tasks, but also increases the revenue of workers.

\subsubsection{Payment Timeliness}

\textbf{Real-time Payment.}
The timeliness of the payment also has an impact on worker motivation. Timely payment of salary can improve the attractiveness of crowdsourcing platforms to workers and increase the stability of crowdsourcing platforms. 
In \citep{Jain-et-al:2016AAMAS}, the authors designed a multiarmed bandit mechanism, CrowdUCB, which is deterministic, regret minimizing, and offers immediate payments to the workers immediately upon completion of an assigned block of tasks. CrowdUCB can quickly learn the quality of workers, maintain better task allocation, and reduce worker turnover by offering immediate payments to workers. Ultimately it achieved the goal of maximizing social welfare.


To realize the goal of maximizing the expected utility while keeping the expected cost within the total budget, the authors proposed a worker arrival model to describe the dynamics of the spatial crowdsourcing system in \citep{Liu-et-al:2020DASFAA}. Subsequently they formulated the problem of Real-time Monetary Incentive for Tasks (MIT) that calculates an appropriate reward for each task in order to maximize the task completion rate in real time. Meanwhile, the authors presented an effective solution for solving the MIT problem with a theoretical effectiveness guarantee, which can greatly reduce the task response time of workers.

\textbf{Delayed Payment.}
Many platforms pay workers with ``delayed payments'' that are employed to protect their own interests when facing the problem that unreliable workers complete tasks after deadline. The Dynamic Price Mechanism (DPM) is proposed in \citep{Chandra-et-al:2015AAAI}, which is not only dominant strategy incentive compatible but also ex-post individually rational. The payment of DPM includes a fixed price module and a dynamic bonus module for ontime completion, which would not be paid to the worker until he has completed all the tasks he has accepted. This mechanism is able to motivate workers to undertake larger contracts and ensures that tasks are completed on time.

\textbf{Milestone Payment.}
Following the same intuition, motivating and retaining workers with payment not only affects the quality of the gathered results, but also affects a crowdsourcer's revenue. It has been found that paying workers based on the milestone works well. Based on this insight, a Hidden Markov Model-based approach \citep{Yin-et-al:2015} was proposed to dynamically determine whether to reward workers with bonuses after a working session in order to maximize the crowdsourcer's expected utility. The key decision here is the timing of distributing bonuses to workers. This approach effectively motivates and retains skilled workers by allocating more bonuses to workers showing better performance.

\subsubsection{Payment Amount}

Monetary incentives not only significantly increase the contribution of workers, but also guarantee task completion rates and task completion times. Nevertheless, from \citep{Bhatti-et-al:2020JSS}, it has been found that the amount of payment is crucial. Large payments can sometimes lead to workers cheating and small payments can occasionally demotivate workers.

Monetary incentives can be used to encourage workers to focus on high valued tasks (e.g., information on traffic conditions at crucial road junctions during rush hour is more important than at other places or times). In \citep{Biswas-et-al:2015}, an incentive mechanism for the generalized setting of an unknown set of workers with non-deterministic availabilities and stochastically rational reporting behaviour was proposed which can produce optimal stochastic solutions on how to disburse payments. Following the ``pay-for-your-own-contribution'' intuition, other approaches motivate high quality workers while discouraging workers with poor performance. In \citep{Radanovic-Faltings:2016}, a boosting scheme - PropeRBoost, was proposed to improve the efficiency of existing incentive schemes. It makes a distinction between low and high quality work to concentration payment to good workers while sanctioning spammers.


In addition, the revenue maximizing price is likely to be limited as a result of supply shortages. In this case, using subsidies to motivate workers is effective. In \citep{Fang-et-al:2017WWW}, the authors proposed an analytic model of agent interactions within a shared platform based on common models for two-sided markets, and captured the asymmetries arising from the interactions between the supply and demand sides. By means of this model, the size of the subsidy can be optimized to maximize the tradeoff between revenue and the cost of the subsidy.

In crowdsourcing markets or systems with various requests (e.g., types of the tasks) and diverse supplies (e.g., skill level and locations of different workers), dynamic pricing strategy is more reasonable, which has been mentioned in Section \ref{Dynamic Salary Distribution}. 
According to the economic principles of monetary incentives, the research on monetary incentives based on game theory has always attracted the attention of researchers. In \citep{Nie-et-al:2018GLOBECOM}, the authors designed a monetary incentive mechanism for the monopoly Crowdsensing Service Providers in order to efficiently recruit sufficient users (workers). The authors modelled the reward process and participation process as a two-stage Stackelberg Game model, and the proposed incentive mechanism can stimulate higher user engagement and greater profits of the Crowdsensing Service Providers, which has been verified through performance evaluation experiments.

However, in more complex situations, such as when there are spatial constraints, time constraints, or limited workers, \methodname{} relying on monetary incentives alone may not guarantee the successful completion of tasks, and designing more comprehensive and effective incentive mechanism remains a significant challenge.

\subsection{Motivating Workers with Mixed Incentives} \label{Motivating Workers with Mixed Incentives}
As crowdsourcing tasks become increasingly complex, and social relationships and network relationships are integrated into crowdsourcing tasks, it is difficult for a single monetary incentive to mobilize the enthusiasm of workers in many complex crowdsourcing tasks. When undertaking crowdsourcing tasks, workers not only pursue money rewards, but also pay attention to the improvement of their own social status. At the same time, in many special crowdsourcing tasks, they also expect to obtain reciprocal benefits. Mixed incentives are designed around worker's \textit{personal status improvement} and \textit{reciprocal benefits} while paying workers rewards.

\subsubsection{Personal Status Improvement}

The personal status in a crowdsourcing collaboration scenario refers to social prestige of the worker, rather than the worker's material wealth or state of health. The improvement of personal status can not only make workers feel psychologically satisfied, but also enhance their influence, bring potential social benefits, and then obtain more monetary benefits. Many researchers design incentive algorithms around workers' social status, most of which focus on reputation incentives.

In \citep{Zhang-et-al:2012INFOCOM}, the authors proposed a game theoretic model based on repeated games to motivate workers to try to work hard. To improve the performance of the noncooperative equilibria, the authors proposed a class of incentive protocols which integrates reputation management mechanisms into the pricing schemes of current crowdsourcing systems. The structure of equilbria is rigorously analyzed and proved, which can help the crowdsourcer design the optimal incentive algorithms and maximize the revenue of the crowdsourcer.

In order to motivate rational workers to complete tasks with high quality, An incentive model based on reputation is designed using repeated game theory in \citep{Rui-et-al:2016JEIT}. Besides, a penalty mechanism is also set up in the incentive model, which will punish malicious workers accordingly. The simulation results show that this mechanism can effectively motivate rational workers to work hard, and the overall performance of the crowdsourcing platform can be greatly improved.
Following the same principle, in \citep{Miao-et-al:2016}, workers' reputation information is used to distinguish their payment in order to smooth short-term fluctuations in performance. The reward system compensates workers based on their reputation, while providing low reputation workers' with chances to rebuild their reputation by working without pay for some time.

The mixed incentive provides both long-term and short-term incentives for workers. In \citep{Wei-et-al:2020Electronics}, the authors devised a decentralized crowdsensing model based on consortium blockchain to ensure the security and privacy of crowdsourcing platform, while designing a hybrid incentive model that takes the reputation and data quality into account to select the appropriate workers, and uses monetary incentive based on bidding price and comprehensive grade to distribute rewards. This model is effective in terms of both short-term incentives and long-term incentives for workers.

\subsubsection{Reciprocal Benefits}

In crowdsourcing tasks, crowdsourcers and workers do not necessarily always have a conflict of interest. For example, in crowdsensing scenarios, crowdsourcers and workers can be mutually beneficial. In the traffic navigation scenario, users (can be regarded as workers) enjoy navigation services and upload traffic status data at the same time. Crowdsourcers provide navigation services and obtain more accurate real-time traffic data from different users.

Meanwhile, workers can also be mutually beneficial. Therefore, some researchers pay attention to the incentive mechanism of crowdsourcers to encourage cooperation among workers in order to obtain greater crowdsourcer benefits.
An incentive mechanism proposed in \citep{Lv-Moscibroda:2015} did not follow the ``pay-for-your-own-contribution" intuition. Instead, the authors studied the problem of motivating crowdsourcing workers and spreading this motivation in incentive networks (e.g., hierarchical crowdsourcing \citep{Nath-Narayanaswamy:2014}). In such networks, a worker's reward depends not only on his own contributions to the tasks, but also in part on the contributions made by his referrals. Such a network-based reward system can be more advantageous than contributions-based schemes when the need for recruiting more workers is important.
Another work in \citep{Gong-et-al:2015TSIPN} proposed an incentive mechanism to stimulate workers' participation, which combined social trust with reciprocity. With the help of this mechanism, arbitrary service requests can be satisfied when adequate social credit can be transferred from users who request more than they are able to provide to those who are able to offer more than they request.


Since crowdsourcing is a social activity, the underlying social network effects can also affect the benefits of users in the collaborative process. Following this principle, the authors proposed a two-stage single-leader-multiple-follower Stackelberg Game model and designed the reward considering the social network effects for motivating the participants in \citep{Nie-et-al:2018GLOBECOM}. Subsequently, the authors proposed the optimal incentive mechanism for discriminatory and uniform rewards respectively in the complete information case, and they obtained the analytical expression for optimal reward. It has been verified by simulation experiments that this mechanism can greatly motivate workers to participate in crowdsourcing work. At the same time, the author investigated the incentive mechanism for the crowdsensing platform with incomplete information on social network effects and proposed an incentive mechanism based on Bayesian Stackelberg Game in \citep{Nie-et-al:2018TWC}, which can help the CSP to achieve greater revenue gain. Further, to be more practical, a general incentive mechanism is proposed in \citep{Nie-et-al:2020TWC} to deal with the coexistence of multiple crowdsensing service providers and users. The authors proposed a Stackelberg Game approach including multiple leaders and multiple followers, where they modelled both the strategies of the CSPs and the social influence of the users into the game process. Experiments on the real-world dataset have verified that the incentive mechanism can improve the participation level of workers and the benefits of crowdsourcers.

Social networks also have an incentivizing effect in the recruitment of crowdsourced workers. Faced with a shortage of workers in mobile crowdsourcing, the authors in \citep{Wang:2020TMC} devised a dynamic incentive mechanism that utilizes social networks to propagate tasks to attract more familiar workers, such as friends. They proposed an improved task-based epidemic model to describe the participants' state changes for task propagation and completion, and the model can offer different rewards according to the real-time status of the worker. The incentive model was experimentally demonstrated to maximise the recruitment and motivation of workers to complete tasks within a financial budget. In \citep{Xu:2022TMC}, the authors proposed an incentive mechanism based on multi-armed bandit, and devised a Stackelberg game to motivating workers in spatial crowdsourcing, which can attract superior workers in the case of unknown qualities of workers and maximize the utilities of participants. The incentive mechanism took workers’ social relations into account to recruit them, meanwhile, the social benefits of workers are modelled into the utility functions and designed in the three-stage Stackelberg Game. 
Nevertheless, different from most of current works about incentive mechanisms where the crowdsourcing platforms strive to offer workers as few rewards as possible, in \citep{Shi:2021TN}, the authors sought to provide extra bonus to induce higher social influences which as a result motivates and enhances worker motivation. They designed the reward mechanism for the dynamic social influence scenario, which ultimately achieves maximum cost saving.

Social networks facilitate the recruitment and motivation of workers, but can also induce worker collusion for profit. To solve this problem, the authors proposed a mechanism based on a truth-detection technology in \citep{Huang:2020TMC}. The mechanism is based on the idea that the correctness of workers' answers to some questions is independently verifiable. 
The authors devised a two-stage Stackelberg Game model, in which the platform optimized the rewards for truth detection in first stage and the workers decided their efforts and result submission strategy in second stage for the purpose of maximizing their own payoffs respectively. The equilibrium analysis demonstrated that worker collusion could be avoided.

\section{Quality Control} \label{Quality Control}
After tasks are delegated to workers, workers are motivated by various incentives to perform them and submit results. The next crucial problem is to ensure the quality of the results. 
The quality of crowdsourcing tasks is influenced by a variety of factors, including the description of the tasks, the delegation of the tasks, the incentives provided, the detection of malicious behaviours and fraudulent results, and workers' profiles, abilities, reputation and collaborative processes. The majority of current \methodname{} research focuses on the problem of improving the quality of the results, and various techniques have emerged to tackle this problem from different perspectives. 


Existing reviews and surveys about quality control in crowdsourcing \citep{Allahbakhsh-et-al:2013IEEEInternetComputing,Zhang-et-al:2016AI,Jin-et-al:2020AI} follow the assumptions that the quality of results and answers is explicitly or implicitly associated with certain aspects of the three entities in crowdsourcing (i.e., crowdsourcers, workers and tasks), and propose to encode these assumptions into quality control mechanisms to effectively control the results quality. 
The vast majority of the algorithms about quality control follow these assumptions:
\begin{enumerate}
\item \textit{Rational Workers}: When workers are intrinsically or extrinsically motivated, their behavioural strategies tend to produce outcomes that are conducive to crowdsourcing quality improvement.
\item \textit{Workers with Quantifiable Skills}: There are variations in workers' skill level, which can be learned based on their past behaviours.

\end{enumerate}

\begin{figure*}[t!]
\centering
\begin{tikzpicture} 
[
    mindmap, every node/.style=concept, concept color=orange, text=white, scale = 0.9, 
    level 1/.append style={level distance=4.1cm, sibling angle=180, font=\large},
    level 2/.append style={level distance=3.5cm, sibling angle=65, font=\small},
    level 3/.append style={level distance=3cm, sibling angle=55, font=\footnotesize}
  ]
  \node{\textbf{\huge{Quality Control}}} [clockwise from=180]
  child [concept color=SkyBlue] {
      node {Worker Quality } [clockwise from=255]
      child { node {Worker Abilities \citep{Rangi-et-al:2018, Dawid:1979, Venanzi-et-al:2014WWW, Imamura-et-al:2018ICML,  Kang:2021ICDM, Guan-et-al:2018AAAI, Yang-et-al:2018} } }
      child { node {Worker Reputation \citep{Yu-et-al:2014AAMAS,Rui-et-al:2016JEIT,Wei-et-al:2020Electronics, Wu-et-al:2012, Oyama-et-al:2013,Venanzi-et-al:2013, Jagabathula-et-al:2014, Xu-Larson:2014, Fu:2022TCC} } } 
      child { node { \small{Worker Behaviours } }[clockwise from=150]
                  child {node[concept,color=PowderBlue,text=white] {\small{Sybil Fraud \citep{Yuan-et-al:2017cikm, Wang:2020SIGKDD, Wang-et-al:2016MobiSys, James:2020TITS} }}} 
                  child {node[concept,color=PowderBlue,text=white] {\small{Collu-sion Fraud \citep{Kuang:2020CN, Li:2021TMC} }}}
                  child {node[concept,color=PowderBlue,text=white, font=\tiny] {\small{Indivi-dual Fraud \citep{Kang:2020TSC, Aydin-et-al:2014, Tarasov-et-al:2014ESWA, Moayedikia:2016ICMLA,  Wu-et-al:2020EUR, Li:2021DSS} }}}
      }
    }
  child [concept color=Orchid] {
      node {Task Quality} [counterclockwise from=270]
      child { node {Instructions Design \citep{Wu-Quinn:2017HCOMP, Gaikwad-et-al:2017CSCW, Manam:2018hcomp, Manam:2019WWW} }}
      child { node {Results Aggregation }[clockwise from=30]
                  child {node[concept,color=Thistle,text=white] {Voting  \citep{Cao:2012VLDB, Aydin-et-al:2014, Li-et-al:2020MM, Wu-et-al:2012} }} 
                  child {node[concept,color=Thistle, text=white] {Sparse Answers \citep{Sun-et-al:2018BIGDATA, Tu-et-al:2018ICDM} }}
                  child {node[concept,color=Thistle, text=white] {Redundant Answers \citep{Li-et-al:2019WWW} }}
      }
      child { node {Quality Assessment }[clockwise from=60]
                  child {node[concept,color=Thistle,text=white] {Expert Assessment \citep{Liu-et-al:2017, Bu:2019IJCS} }} 
                  child {node[concept,color=Thistle, text=white] {Worker Assessment \citep{Wang-et-al:2017, Baba-et-al:2013SIGKDD, Yang-et-al:2018} }}
      }
      child { node { \small{Quality Improvement} }[clockwise from=100]
                  child {node[concept,color=Thistle,text=white] {\small{Higher Quality Results \citep{Mo:2013KDD, Zhang-et-al:2015CIKM, Xu-et-al:2021IS} }}} 
                  child {node[concept,color=Thistle,text=white] {\small{Lower Latency \citep{Liu-et-al:2020DASFAA, Manam:2019WWW,  Cheng-et-al:2018TKDE} }}}
      }
    };
\end{tikzpicture}
	\caption{The \methodname{} taxonomy on quality control in crowdsourcing systems.} \label{fig:QC}
\end{figure*}
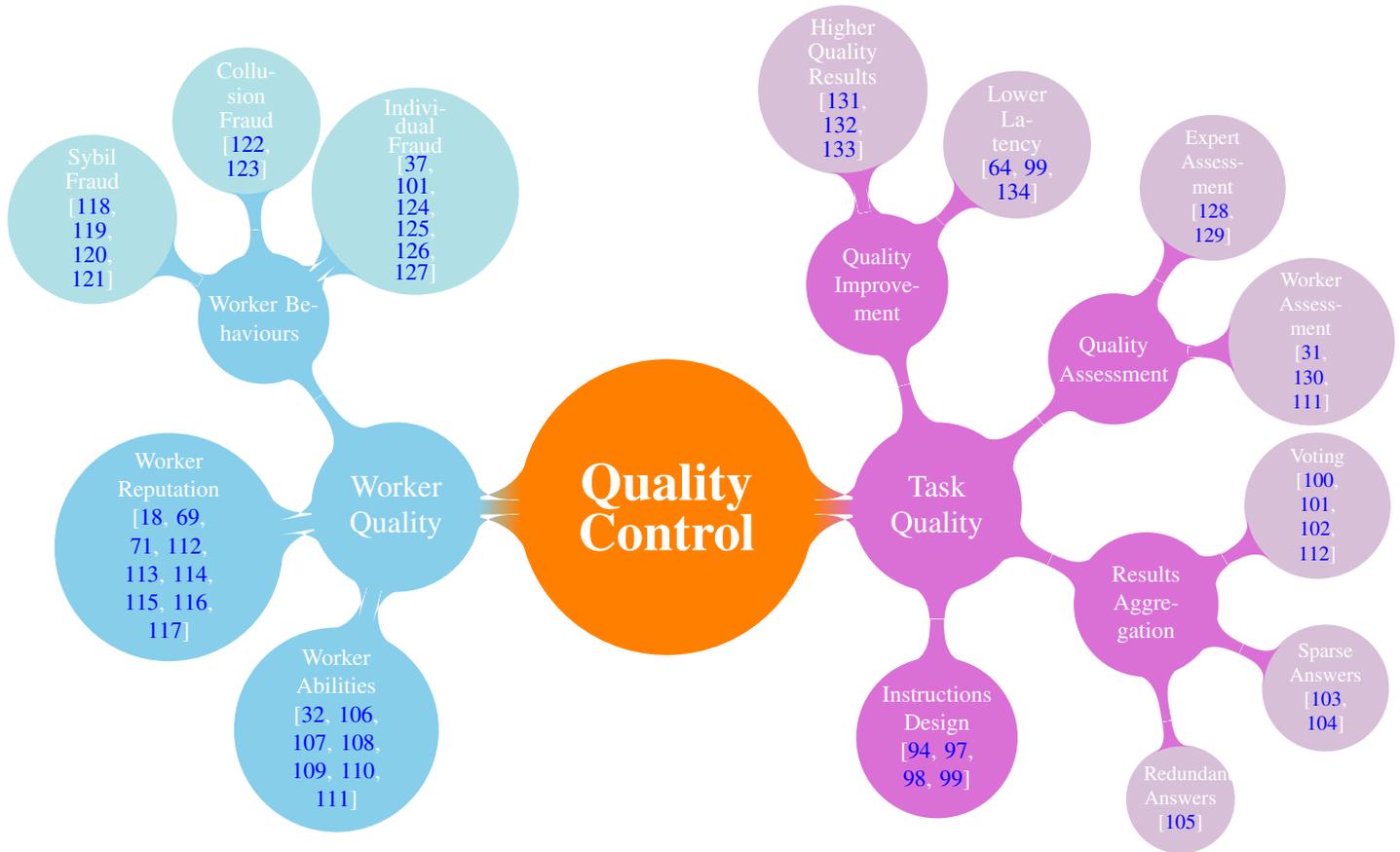

Without quality control of crowdsourcing, the results and task outputs collected from the crowd are inherently confusing and not sufficiently accurate. Different from the existing review studies of quality control algorithms, we summarise and listed existing quality control algorithms that can improve crowdsourcing results from two perspectives, namely task quality control (Section \ref{Task Quality Control}) and worker quality control (Section \ref{Worker Quality Control}), which is in line with the idea that ensuring worker quality and task quality is the important means of enabling system intelligence \citep{Chai:2017IJCS}. Figure \ref{fig:QC} illustrates our proposed classification framework for crowdsourcing quality control algorithms.

\subsection{Task Quality Control} \label{Task Quality Control}

A well-designed crowdsourcing task, especially a clear task instruction, is a prime requirement for success. After the workers have completed their work according to the task requirements and submitted their respective results to the crowdsourcer, the crowdsourcer aggregates the results or answers of the different workers into a target output and evaluates the results. Thus, from the perspective of a task, in addition to the task delegation algorithms mentioned in Section \ref{Task Delegation}, the algorithms for task quality control focus on task instructions design, results aggregation, quality assessment and quality improvement, which are integrated throughout the process of task quality control.

\subsubsection{Task Instructions Design}

Task quality control should start at the source of the crowdsourcing process. In other words, firstly, we need to ensure that effective task instruction design is in place. The task instructions of a crowdsourcing task not only affect the efficiency of the task execution, but also have a significant impact on the task outcomes. Besides, task instruction quality has been widely presumed to influence result quality and latency time. There are best practice guidelines written by experienced crowdsourcers on how to design task instructions. A study on mTurk \citep{Wu-Quinn:2017HCOMP} found that most tasks follow the best practice guidelines, but many did not explain the acceptance criteria clearly. However, workers appear to view these tasks favourably. In addition, although having more detailed task instructions can help improve result quality, it negatively affects worker uptake as they have to read through and understand more information about the tasks. This, in turn, increases the latency time. 
In addition, complex task interface and description will affect the workers' perception of the task, thus affecting the quality of the results \citep{Kittur-et-al:2013CSCW}.

In \citep{Baba-et-al:2013AAAI}, the authors proposed an automatic approach to prevent crowdsourcers from posting illegal or objectionable tasks. This is achieved through a second level of crowdsourcing which engages workers to monitor the task instructions to identify such contents. In crowdtesting tasks which require workers to give feedbacks to crowdsourcers' products or services, it has been found that under anonymous conditions, workers provide significantly more specific criticisms and praises. These feedbacks are also rated as more useful by the crowdsourcers. 


Existing models and algorithms for controlling crowdsourcing quality from the perspective of task design are mainly based on crowdsourcer-worker interaction to quickly identify or locate confusing task instructions, which are then explained in detail or modified by crowdsourcers. In \citep{Gaikwad-et-al:2017CSCW}, the authors proposed Daemo, a mechanism for proactively identifying ambiguities in task instructions. By posting several task instances to workers and obtaining their feedback, it can improve the task based on the feedback. The experimental results showed that the task results are significantly better with the use of improved tasks. In order to identify and tackle the unclear task instruction problems in a more general way, rather than just solving the problem of vague descriptions in a few instance tasks, the authors designed a method to improve the task quality for unclear or ambiguous task instructions in \citep{Manam:2018hcomp}. When the requester is active, the proposed approach improved the task instructions through interaction between the worker and the requester. When the requester is inactive, the authors designed an iterative workflow that allowed a worker to collaborate and improve answers from other workers. Furthermore, based on the experience in the previous study \citep{Wu-Quinn:2017HCOMP}, a system developed in \citep{Manam:2019WWW}, TaskMate, can help requesters create high quality instructions in a short time by recruiting a reliable set of crowd workers, ultimately ensuring consistent interpretations by crowd workers for the same task and reducing wasted work time.

\subsubsection{Task Results Aggregation}

Crowdsourcing is the product of crowd intelligence, and aggregating the wisdom of the crowd to the greatest extent can improve the quality of the results. Hence, many researchers have proposed algorithms to control the quality of crowdsourcing from the perspective of worker outcome aggregation. Voting strategies are commonly adopted to infer the results of the task, such as Majority Voting \citep{Cao:2012VLDB}, Weighted Majority Voting  \citep{Aydin-et-al:2014}, Probabilistic Graphical Models \citep{Li-et-al:2020MM}, etc. In \citep{Cao:2012VLDB}, the authors proposed two algorithms and devised an bounding technique which greatly reduce the error rate of answers aggregated by Majority Voting. Subsequently, another team of researchers proposed a method to aggregate answers for multiple choice tasks in \citep{Aydin-et-al:2014}, which combined the weights of worker confidence and worker reputation. The experimental results showed that the proposed method can find the correct answers for those harder questions. Probabilistic graphical model is another way to efficiently aggregate results. In \citep{Li-et-al:2020MM}, the authors proposed a probabilistic graphical annotation model to infer the underlying ground truth while modelling the behaviours of different annotators, which can help identify the probability of annotators producing noisy labels during the test.


When there are not enough workers or when workers only give answers to a small number of questions due to capacity and time constraints, the data collected by the crowdsourcers is often sparse, which can significantly affect the downstream tasks. Matrix factorization algorithms are typically used to solve data sparsity problems. In \citep{Sun-et-al:2018BIGDATA}, a matrix factorization algorithm under local differential privacy is designed to improve the quality of answers, which can address the trade-off between privacy and utility. The proposed matrix factorization algorithm can satisfy local differential privacy requirements while reducing the error of truth inference, which is also effective when workers' answers are very sparse. Similarly, another algorithm proposed in \citep{Tu-et-al:2018ICDM} employed the principle of matrix factorization to aggregate the labels of multi-label samples collected via crowdsourcing. It jointly decomposed the sample-label association matrices obtained from different workers into the product of individual low-rank matrices and a shared low-rank matrix, and assigns different weights to their answer data matrices for selective integration.

Conversely, there are sometimes a large number of redundant worker answers, which can reduce the quality of the crowdsourcing results. In \citep{Li-et-al:2019WWW}, the authors proposed a generative Bayesian model based on a normal likelihood and conjugate inverse-Gamma prior, which modelled a discrete problem as a continuous regression problem. Experimental results showed that the proposed model achieved the highest average accuracy on 10 real data sets of different sizes and varying degrees of label redundancy from several different application domains.

\subsubsection{Task Quality Assessment}
Quality assessment approaches for estimating the quality of results are mainly distinguished by the number of reviewers involved in assessing the same task: 1) a single reviewer (mostly expert) by rating and 2) multiple reviewers by voting. Involving experts in quality management is safe yet very expensive and laborious. In \citep{Liu-et-al:2017}, the authors proposed a semi-supervised learning algorithm which selects the most informative results based on a hybrid uncertainty model, and propagate the expert labels to similar instances using a loss-driven max-margin majority voting algorithm. This approach efficiently reduces the experts' effort up to 60\%.

An intuitive idea to improve the quality of results without involving experts is to assign the same task to multiple workers redundantly, and then aggregate the results to estimate the ground truth via majority voting \citep{Tian-et-al:2015}. However, such redundant task assignment techniques are limited by the underlying assumption that task complexity and worker capability are similar, which is not always true. In \citep{Sunahase-et-al:2017}, the authors adopted a two-stage procedure consisting of a creation stage and an evaluation stage. In the creation stage, workers (called creators) first create artifacts. In the evaluation stage, another set of workers (called evaluators) compare a pair of artifacts and vote for one of them. This quality estimation method is based on an extension of Kleinberg’s HITS algorithm, which accounts for the ability of both the evaluators and the creators. 
Another unsupervised statistical quality estimation algorithm proposed in \citep{Baba-et-al:2013SIGKDD} also employed similar two-stage procedure. Different from \citep{Sunahase-et-al:2017}, the authors exploited the notion of mutual evaluation between different workers and modelled the ability of each workers, then a probabilistic generative model based on the graded response model was developed. Experimental results implied that the proposed model can improve the quality of results while keeping lower costs.
The authors in \citep{Bu:2019IJCS} paid more attention to the quality of the multiple-step classification tasks and proposed a fancy model to capture the workflow information for both single-question and multiple-question classification tasks, which is also a novel exploration.

In addition, deep learning has been employed to assess the quality of results aggregated from multiple workers with unknown expertise. The approach is general and robust to noise and data sparsity. In \citep{Yang-et-al:2018}, the authors proposed a deep Bayesian learning framework to optimize the estimations for both annotated results and workers' expertise for the purpose of minimizing the number of required annotation instances and workers. This approach is a generalized version of the former uncertainty based models \citep{Wang-et-al:2017, Liu-et-al:2017} with increased computation cost. 

Crowdsourcing not only relies on workers to complete complex tasks, but with the rapid development of IoT technology and smart devices, mobile smart devices with powerful sensing capabilities are also playing an important role as task performers, especially in mobile crowdsensing tasks. To ensure the reliability of sensing data and to meet the requirements of collaborative communication between collaborators, the authors proposed a social team crowdsourcing framework, TAQ-Crowd in \citep{Liu:2021CN}. Within TAQ-Crowd, trusted relationships between nodes are modelled for sensing data quality assessment, while a tree routing network is constructed to guarantee communication quality based on the classical Traveling Salesman Problem. Extensive simulation showed TAQ-Crowd framework is effective in quality estimation.

\subsubsection{Task Quality Improvement}

Quality assessment is not the fundamental goal of crowdsourcing, while improving the quality of crowdsourcing results is the ultimate pursuit of the crowdsourcer and the task requester. In addition to algorithms such as optimising task design and aggregating results from a task perspective to improve crowdsourcing quality, many researchers have used the implicit knowledge of multiple tasks to achieve collaborative improvement. In \citep{Mo:2013KDD}, Transfer Learning is integrated into the hierarchical Bayesian model. By considering the overlapping workers as a bridge, the proposed model can solve the data-sparsity problem by exploiting knowledge from related tasks, which can significantly reduce the cost of the crowdsourcers while improving the quality of the tasks.

Improving the quality of crowdsourced data labels and thus the accuracy and precision of results is a constant pursuit for crowdsourcers and task requesters. In \citep{Zhang-et-al:2015CIKM}, the authors developed a new framework to improve the label quality by introducing noise correction techniques, and proposed a novel adaptive voting noise correction algorithm (AVNC), which can identify and correct noise with the assistance of the estimated label quality delivered by ground truth inference. The experimental results showed AVNC can significantly improve label quality when using a variety of different inference algorithms. Another work in \citep{Xu-et-al:2021IS} proposed a noise correction approach based on cross-entropy, which makes use of the entropy of the label distribution to filter noisy instances. The cross-entropy values between possible true class probability distribution and predicted class probability distribution is then used to improve the quality of the instances.


There is a trade-off between latency and quality, in other words, improving the quality of crowdsourcing cannot ignore latency issues. The work in \citep{Manam:2019WWW} focused on improving the quality of task instructions to save workers' task execution time and the work in \citep{Liu-et-al:2020DASFAA} paid more attention to motivating workers to reduce the task response time of them. Generally speaking, setting longer completion times when posting tasks to workers can improve the quality of results, but will incur higher task costs and poor timeliness. In \citep{Cheng-et-al:2018TKDE}, the authors proposed the FROG framework to balance low latency with high accuracy, which includes two modules: task scheduler module, in which two different dimensional scheduling algorithms are devised to ensure reliability of task execution and reduce task latency, and notification module, which chooses to notify the workers who are more likely to accept tasks as task performers by means of kernel density estimation and further guarantees task reliability.
Based on the intuition that more reliable workers are likely to perform difficult tasks better, the authors proposed a kernel ridge regression function to quantify the difficulty of tasks in \citep{Tu:2020TKDD} and assigned difficult tasks to more reliable workers, which ultimately ensures the quality of task completion.

Quality assessment not only occurs after task submission, but also a small number of studies focus on the task assignment phase. For example, Zheng et al.\citep{zheng:2015SIGMOD} incorporated evaluation metrics into task assignment to ensure task quality and worker benefits

\subsection{Worker Quality Control} \label{Worker Quality Control}

Another perspective for controlling the quality of crowdsourcing is from the entity of the worker. Choosing honest, rational and high-quality workers to participate in a task often leads to better results. The factors that influence worker quality mainly include the worker's abilities, reputation and irrational behaviour, which can be derived by means of worker individual assessments, such as qualification tests\citep{Heer:2010CHI}, personality tests\citep{Kazai:2012CIKM} and behavioural data data mining.

\subsubsection{Worker Abilities}

An intuitive idea for controlling quality is to select workers with high ability to complete tasks and to eliminate those with low ability. Worker ability is the first studied and widely recognised factor influencing the quality of crowdsourcing work. The effect is assumed to be determined by the abilities or skill level of each worker.

In \citep{Dawid:1979}, the authors developed a model to estimate the ability of worker, which is the first study to consider worker abilities. The model adopted the expectation maximization algorithm to provide maximum likelihood estimation of the responses of all workers. However, this model tends to fall into the dilemma that the confusion matrix is too sparse when there are few workers or worker answers. To address this problem, the authors employed Bayesian hierarchical model in \citep{Venanzi-et-al:2014WWW} to cluster the workers to take full advantage of the similarities between workers, ultimately smoothing the worker confusion matrix. In a similar situation, another study \citep{Imamura-et-al:2018ICML} directly replaced worker confusion matrix with the corresponding group confusion matrix. Besides, they utilized log-likelihood function rather than using Bayes rule. Both studies were able to further accurately measure worker abilities and improve the quality of crowdsourcing. Similar models and methods for quantifying worker skills are presented in \citep{Tang:2020IJCAI,Li:2021DSS}.

In \citep{Rangi-et-al:2018}, the authors formalized the notion of workers' ability and proposed an online strategy to estimate it in terms of value contributions of the workers. Meanwhile, the authors formalised the task delegation problem as a multi-armed bandit set-up, and used the B-KUBE algorithm to tackle it. Combined with the proposed online estimation algorithm for workers' ability, B-KUBE delegated tasks to the workers with high abilities, ultimately achieving a lower misclassification rate and completing more tasks.
In \citep{Kang:2021ICDM}, the authors designed a framework based on self-paced learning methodology to stimulate workers' learning ability, and proposed a task delegation strategy based on benefit maximization concept. The framework is able to dynamically assess worker capability and task difficulty to ensure the quality of tasks by pushing difficult tasks to groups of high-quality workers.


Deep learning methods are also frequently employed to model and learn the abilities of workers, with the goal of improving the quality of crowdsourcing. The generic Bayesian framework incorporating deep learning proposed in \citep{Yang-et-al:2018} leveraged the low-rank structure to learn the expertise of individual worker, which represents different skill levels of crowdsourced workers. The framework is experimentally shown to accurately learn the expertise of annotators, infer true labels, and effectively reduce the amount of annotations in the model training, which can reduce the cost of crowdsourcing while guaranteeing quality. In \citep{Guan-et-al:2018AAAI}, the authors utilized a share neural networks and learned averaging weights to predict the ability of each experts (workers) independently, which acts as a weight on majority vote on the label prediction task. This method can be helpful to exploit the various types of strengths of different workers in a crowdsourcing task.

\subsubsection{Worker Reputation}
In a crowdsourcing campaign, workers who focus on their influence and reputation will keep producing high quality work until they no longer care about their own influence and the attention they attract. To a certain extent, workers who gain higher reputation and higher social influence are not only more incentivised to continue to participate in crowdsourcing, but their work tends to be of higher quality because of the higher reputation as a recognition of their work by other workers and crowdsourcers. Over the past decade, several crowdsourcing task delegation algorithms and incentive mechanisms that consider worker reputation, such as \citep{Yu-et-al:2014AAMAS,Rui-et-al:2016JEIT,Wei-et-al:2020Electronics} have been shown to be effective in improving platform performance while ensuring task quality.

Workers' performance track records, often summarized with various reputation modelling techniques, can be used as a guide to improve future quality of results. In \citep{Wu-et-al:2012}, the authors aim to improve the quality of speech labels from crowd workers. The proposed statistical model, Sembler, combines majority voting, reputation modelling for annotators, and linguistic contexts in order to guarantee the quality of crowd labeling from non-experts. 


Nevertheless, as crowdsourcing tasks are often small and assigned to multiple workers redundantly, it is costly in practice for crowdsourcers to provide feedbacks on the quality of all the results obtained. Thus, reputation-based approaches need to adapt to the lack of direct feedbacks in order to operate in crowdsourcing systems. In \citep{Oyama-et-al:2013,Venanzi-et-al:2013}, the authors incorporated a trust model into a fusion method which merges task results based on the trust parameters. They further provided an inference algorithm that jointly computes the aggregated result and the workers' individual trustworthiness based on the maximum likelihood framework to address this challenge. More broadly, in \citep{Jagabathula-et-al:2014}, a deterministic adversarial strategy was proposed which incorporates disagreement as a penalty into the reputation model which is robust when inferring ground truth from noisy results.



While worker reputation can be used to aggregate the results for simple tasks, in winner-takes-all contest-based crowdsourcing (e.g., design contests in 99designs.com), worker reputation is used differently. In such contests, workers whose entries are not selected receive no reward. Thus, social welfare can be poor due to wasted effort. In \citep{Xu-Larson:2014}, the authors proposed a discrete choice model to capture workers' output qualities and designed a mechanism to filter out low-expertise workers before they are asked to produce a solution to enter the contest. In this way, the efficiency of such contests can be improved.

The recent rapid development of blockchain technology can also be used for crowdsourcing quality control, especially for managing workers' behaviours and reputations. In \citep{Fu:2022TCC}, the authors proposed a blockchain-based crowdsourcing framework with reputation and incentives, and the authors utilised reputation as a measure of employee trustworthiness, which is derived via a novel subjective logic model. Combined with incentives based on contract theory, this framework can detect malicious workers and defend against conspiracy fraud to some extent.

\subsubsection{Worker Behaviours}


Some selfish workers may submit random answers and commit fraud against the crowdsourcing platform in order to gain more monetary incentives, which will lead to a reduction in the quality of results and ultimately a great loss to the crowdsourcers and crowdsourcing platforms. Many studies have consequently been devoted to detecting malicious and unreliable workers involved in crowdsourcing activities and their fraudulent behaviours. Although incentives such as delayed payments \citep{Chandra-et-al:2015AAAI} can discourage fraudulent workers and reduce losses to the platform and task requesters, they do not prevent fraudulent workers from appearing. Therefore, it is essential to detect fraudulent workers.


\textbf{Sybil Fraud.}
A common category of fraud in crowdsourcing is the generation of a large number of sybil workers competing for tasks. The sybil workers who are fraudulent and low-quality participate in the task competition process for profit, affecting the quality of the task and disrupting the crowdsourcing market, which is attributed to sybil attacks in crowdsourcing. A sybil attacker is likely to manipulate multiple sybil worker accounts to share random answers of unknown correctness for each task, so that the responsible high-quality and independent workers on the same task may be rejected in the competition.

Many studies have been conducted on the detection and defenses of Sybil attacks. 
In \citep{Yuan-et-al:2017cikm}, the authors proposed a cluster-based sybil defense framework and devised a cluster-based Sybil attack detection algorithm. By defining the similarity of workers and designing a worker similarity function, the algorithm is able to cluster workers and furthermore detect Sybil groups through some gold standard questions. Experimental results showed that the proposed algorithm is effective against multiple Sybil attackers and can accurately detect Sybil workers online with low cost.
However, the Sybil attacker may identify gold standard tasks and disguise themselves to evade detection. In \citep{Wang:2020SIGKDD}, the authors proposed a probabilistic task delegation approach to strategically assign golden tasks while camouflaging them from the Sybil attackers. Meanshile, they designed an online framework to solve the shortage problem of golden tasks, which achieved a high aggregation accuracy even in the face of the strategic attack from sybil workers.

In addition, multiple types of devices in spatial crowdsourcing services can be a source of fraud for sybil attacks. The authors identified and studied a range of attacks on crowdsourced map services in \citep{Wang-et-al:2016MobiSys}, and then proposed diverse techniques to support the construction of the proximity graphs, which are used to perform the detection of Sybil devices.
In \citep{James:2020TITS}, the authors designed a deep generative model, Bayesian Recurrent Autoencoder (BRAE), to capture the sequential characteristics about vehicular trajectories, then proposed a deep learning sybil attack identification algorithm to tackle the sybil attack problem in crowdsourced navigation. BARE implemented a multivariate random variables representation for trajectories by means of Bayesian recurrent neural network as the Trajectory Encoder and used self-supervised learning on historical track data to learn potential representations. 

\textbf{Collusion Fraud.}
The second category of fraud is worker collusion fraud. There are certain similarities between worker collusion fraud and sybil fraud, the difference being that the subject of the benefit becomes multiple workers and the behaviour of these colluded workers remains similar. In \citep{Kuang:2020CN}, the authors summarized three collusion patterns of spam workers and formulated the spammer detection problem as a node classification task on the defined heterogeneous network. They learned the worker representation by means of a heterogeneous network embedding model and utilized the one-class Support Vector Machines algorithm to classify workers according to the learned worker representation, which ultimately enabled the detection of various types of collusive fraud workers. In \citep{Li:2021TMC}, the authors proposed a misreport- and collusion-proof crowdsourcing mechanism based on game theory, which aims to encourage workers to be as honest as possible in reporting the quality of the results of the tasks submitted. They designed a two-worker incentive mechanism that satisfies the incentive compatibility and the collusion-proof constraints, and further extended it to multiple workers. Extensive simulation results showed the mechanism is effective in reducing the risk of worker collusion fraud when verifying the results of crowdsourcing tasks.


\textbf{Individual Fraud.}
Another type of crowdsourcing related to low quality workers is the occurrence of random or unreliable behaviour due to selfish or lazy workers, producing unreliable results, which belongs to Individual Fraud and also requires timely detection. A practical MAB-based approach is proposed in \citep{Tarasov-et-al:2014ESWA} for the dynamic estimation of worker reliability in regression, which requires no previous knowledge about the task and can be easily employed for a wide range of binary and multi-classification tasks. In \citep{Moayedikia:2016ICMLA}, the authors proposed a dynamic reliability estimation algorithm based on bee colony algorithm (REBECO), which leverages a two-stage reliability estimation process for workers: exploration phase for allocating tasks and exploitation phase for measuring workers' reliability. After that, the workers are divided into two separate groups of reliable and unreliable. At the same time, they devised two algorithms to deal with unreliable workers: impatient variation algorithm that removes unreliable workers instantly and patient variation of REBECO algorithm that preserved unreliable workers and reduced the payment to them.

It is important to ensure the reliability of workers. In addition to worker reputation computed from their track records, their self-reported confidence can also be a useful source of information for quality control. In \citep{Sakurai-et-al:2013}, the authors designed a method which allows workers to declare their confidence indirectly by choosing from a set of reward plans each corresponding to a different confidence level. In this way, the approach addresses the problem of workers being over-confident and/or untruthful. Similarly, in \citep{Aydin-et-al:2014}, the authors combined machine learning with workers' reliability model derived form their self-reported confidence to improve the aggregation accuracy for multiple-choice questions. In \citep{Kang:2020TSC}, the authors proposed a multi-armed bandit framework that learns a worker's preferences and reliability over time in the absence of the prior knowledge about the worker. In \citep{Shi:2021INFOCOM}, the authors designed a truth inference algorithm to estimate the reliability of workers and dynamically updated workers' reliability. Meanwhile, they designed an online task assignment mechanism with the maximum reduced ambiguity principle to maximize the quality of task allocation. 
Besides, deep learning techniques are also worth considering. In \citep{Wu-et-al:2020EUR}, the authors constructed a back-propagation CNN-like deep neural network with a crowdsourcing layer that identifies unreliable workers to minimize task failure.

In order to ensure a higher degree of truthfulness of both answers and worker profiles in mobile crowdsourcing, the authors devised incentives to motivate workers to submit true task answers and profiles in \citep{Xiao-et-al:2022TMC}. The authors analysed the sufficient and necessary conditions for answer truthfulness and task truthfulness separately and used these conditions to construct an incentive optimisation problem whose solution is the reward that should be paid to workers for performing honest behaviours. A Bayesian Nash equilibrium is also guaranteed through an analytical proof, which is demonstrated in a practical experiments demonstrate the effectiveness of the mechanism. 
In addition, the connectivity of social relationships graph can be used to identify whether an worker is trustworthy or not. The authors focused on building a framework to learn workers' preferences, skills and willingness to participate in crowdsourcing based on social networks in \citep{Li:2021DSS}, and devised a recommendation method to select proper workers as the task candidate performers according to calculated suitability score to the task.

The starting point of the quality control algorithms, both from the task perspective and from the worker's perspective, is to advance the implementation of the tasks, improve the quality of the results and promote the sustainability of the crowdsourcing platforms. The quality control algorithms are inextricably linked to the task delegation algorithms, as well as the incentive mechanism algorithms, which are summarized in the previous sections. Together, these three types of algorithms form the main algorithmic framework for a closed-loop crowdsourcing process.

\section{Future Research Directions} \label{Future Research Directions} 
With myriad techniques proposed for task delegation, motivating workers and quality control, much work is required to make \methodname{} widely acceptable for practical applications. For example, how to ensure fair competition among crowdsourcers to reduce the occurrence of industry monopolies, how to reduce malicious fraud by workers to improve the quality of tasks, and how to protect the privacy of crowdsourcers and workers, etc. In addition to the aforementioned potential problems of improvement, we highlight five promising future research directions in \methodname{}.

\subsection{Effective Task Assignment and Delegation}

Current task allocation research mostly centres around offline task assignment, which can seldom be used in practical real-world applications (e.g., Uber) requiring online task assignment and delegation. How to achieve optimal dynamic assignment of tasks to ensure the maximum benefit of the platform in response to the constant arrival of instant tasks remains a significant challenge. Meanwhile, there is still a gap in efficient execution or delegation decision algorithms for instant tasks.



\subsection{Federated Crowdsourcing}
With the increasingly concerns on data security, privacy protection in crowdsourcing research has became extremely important. 
Prior researches usually inject noise to participants’ sensitive data through techniques such as differential privacy. 
However, this method would inevitably lead to quality loss to crowdsourcing. 
Recently, the federated learning (FL) paradigm has been proposed which enables participants to collaboratively train models without exposing their raw data, which could realize privacy-preserving model training with little loss (or even no loss) of model performance. \citep{wang2020federated,Shi-et-al:2023ICWS} have presented federated crowdsourcing frameworks by incorporating federated learning techniques into four stages of crowdsourcing, such as task creation, task assignment, task execution and data integration. Some specific research challenges and opportunities of federated crowdsourcing still needed to be addressed such as energy consumption, network connection and transfer learning in federated crowdsourcing.

\subsection{Regulated and Explainable Process Control}
Currently, most recommendation-based approaches assume that the suggestions would be adhered to by all. However, such an assumption might not be valid in reality. More research effort for understanding human compliance patterns and how \methodname{} approaches can adapt to non-compliance is needed. In addition, techniques such as explainable AI \citep{Zeng-et-al:2018} and persuasive design \citep{Kang-et-al:2015} can be explored to help \methodname{} techniques build trust with users to improve compliance. Nevertheless, much work is still needed to determine the optimal amount of information to be included in such explanations and persuasions in order to engender trust while not overwhelming the users. Ethical and governance issues \citep{Yu-et-al:2018IJCAI,Zhang-et-al:2022IJCS,Zhang-Yu:2022IJCS} in the context of \methodname{} also need to be studied in order to make the technology effective, fair and privacy-preserving \citep{Yang-et-al:2019TIST,Yuan:2020TIFS,Li:2022TKDE}.

\subsection{Reputation Building and Management of Platforms}
With the emergence of more and more crowdsourcing platforms, competition among platforms is becoming more prominent. When faced with similar tasks on different platforms, workers often choose the platform with the better reputation to participate in the task, which poses new challenges for platforms to recruit workers. Hence, it is a worthwhile direction to explore in depth how crowdsourcing platform can recruit more workers, build a good reputation and achieve a trade-off between reputation management and cost savings on a limited budget.

\subsection{Transparent, Secure and Trusted Crowdsourcing}
Trusted security is becoming an indispensable guarantee in the process of AI-enabled crowdsourcing. Current \methodname{} approaches generally give more control to AI. More transparency and control shall be provided to the humans in the loop. Research on interactive crowdsourcing where AI and humans influence each other and jointly control the crowdsourcing process may be a promising future direction. Crowdsourcing, as a social activity, still faces the dilemmas of traditional social collaboration when AI empowers crowdsourcing, such as the user information security issues, trustworthiness of the collaboration processes, whether the interactions are transparent and tamper-proof in the crowdsourcing, which are all huge challenges for the future and may be able to be solved using the evolving and maturing trusted AI technologies, such as blockchain technology.

\subsection{Open Platforms for Testing and Data Sharing}
Last but not least, current \methodname{} research mostly rely on computer simulations for evaluation. This is partially due to the lack of access to real-world crowdsourcing platforms. The research community shall consider pooling resources to build a testbedding platform for \methodname{} research with access to human participants. Such an open approach can help the field make more rapid and measurable advances in the future.

\section*{CRediT authorship contribution statement }
\textbf{Shipeng Wang:} Paper collection and analysis, conceptualization, manuscript writing, drawing of figures. 
\textbf{Qingzhong Li:} Conceptualization, manuscript review and proofreading.
\textbf{Lizhen Cui:} Conceptualization, manuscript review.
\textbf{Zhongmin Yan:} Paper collection and analysis, revision of figures. 
\textbf{Yonghui Xu:} Paper collection and analysis.
\textbf{Zhuan Shi:} Paper collection and analysis, manuscript writing.
\textbf{Xinping Min:} Paper collection and manuscript review.
\textbf{Zhiqi Shen:}Manuscript review and proofreading. 
\textbf{Han Yu:} Conceptualization, manuscript review, writing and proofreading.

\section*{Acknowledgments}
Qingzhong Li and Han Yu are the corresponding authors. This research is supported, in part, by National Key R\&D Program of China No. 2021YFF0900800; Shandong Provincial Major Scientific and Technological Innovation Project NO.2021CXGC010108; the National Research Foundation Singapore and DSO National Laboratories under the AI Singapore Programme (AISG Award No: AISG2-RP-2020-019); the Joint SDU-NTU Centre for AI Research (C-FAIR); the RIE 2020 Advanced Manufacturing and Engineering (AME) Programmatic Fund (No. A20G8b0102), Singapore; and the Nanyang Assistant Professorship (NAP), Nanyang Technological University.






\bibliographystyle{elsarticle-num-names}
\bibliography{KBS.bib}





\end{document}